\newcommand{\nn}{\nonumber}

\newcommand{\cH}{{\cal H}} 
\newcommand{\cC}{{\cal C}} 
\newcommand{\cN}{{\cal N}} 
\newcommand{\cV}{{\cal V}} 
\newcommand{\cD}{{\cal D}} %
\newcommand{\cE}{{\cal E}} %
\newcommand{\cZ}{{\cal Z}} 

\newcommand{\bR}{{\bf R}} 
\newcommand{\bX}{{\bf X}} 
\newcommand{\bZ}{{\bf Z}} 
\newcommand{\bN}{{\bf N}} 
\newcommand{\Bket}[2]{{|\,B:#1,#2\rangle\!\rangle}}
\newcommand{\Bbra}[2]{{\langle\!\langle B:#1,#2\,|}}

\newcommand{\Ckets}[2]{{|\,C:#1,#2\rangle\!\rangle}}

\newcommand{\tn}{{\overline{n}}}
\newcommand{\tm}{{\overline{m}}}
\newcommand{\tp}{{\overline{p}}}
\newcommand{\tq}{{\overline{q}}}
\newcommand{\tk}{{\overline{k}}}
\newcommand{\tl}{{\overline{l}}}
\newcommand{\tI}{{\overline{I}}}
\newcommand{\tJ}{{\overline{J}}}
\newcommand{\tK}{{\overline{K}}}

\newcommand{\cycl}{{\cal T}}
\newcommand{\cyclb}{{\overline{\cal T}}}
\newcommand{\inv}{{\cal S}}

\newcommand{\be}[1]{{\bf e}\!\left[#1\right]}

\newcommand{\mod}[1]{{(\mbox{mod}\,\,\, #1)}}

\newcommand{\diag}{{\rm diag}}

\documentclass[12pt]{article}
\usepackage{epsf}
\usepackage{amssymb}

\catcode`\@=11
\@addtoreset{equation}{section}

\catcode`@=12

\begin{document}

\renewcommand{\thefootnote}{\fnsymbol{footnote}}
\font\csc=cmcsc10 scaled\magstep1
{\baselineskip=14pt
 \rightline{
 \vbox{\hbox{UT-968}
       \hbox{December 2001}
}}}

\vfill
\begin{center}
{\Large\bf
Open Superstring on Symmetric Product
}

\vfill

{\csc Hiroyuki Fuji}\footnote{
      e-mail address : fuji@hep-th.phys.s.u-tokyo.ac.jp}\\
\vskip.1in

{\baselineskip=15pt
\vskip.1in
  Department of Physics,  Faculty of Science\\
  Tokyo University\\
  Bunkyo-ku, Hongo 7-3-1, Tokyo 113-0033, Japan
\vskip.1in
}

\end{center}
\vfill

\begin{abstract}
{
The string theory on symmetric product describes the second-quantized
string theory.
The development for the bosonic open string was discussed in the
previous work.\cite{Matsuo-Fuji}
In this paper, we consider the open superstring theory on the symmetric
product and examine the nature of the second quantization.
The fermionic partition functions are obtained from
the consistent fermionic extension of the twisted boundary conditions
for the non-abelian orbifold, and they can be interpreted in terms of the
long string language naturally.
In the closed string sector, the boundary/cross-cap states
are also constructed.
These boundary states are classified into three types in terms of the
long string language, and explain the change of the topology of
the world-sheet.
To obtain the anomaly-free theory,
the dilaton tadpole must be cancelled.
This condition gives $SO(32)$ Chan-Paton group
as ordinary superstring theory.
}
\end{abstract}
\vfill

\setcounter{footnote}{0}
\renewcommand{\thefootnote}{\arabic{footnote}}
\newpage

\section{Introduction}
The second quantization of the string is one of the most important
issue in the string theory.
There has been considerable works for
the second-quantization of the string theoy.\cite{Kaku-Kikkawa,WITTEN,HIKKO}
Especially the extension for the superstring
theory\cite{Green,Berkovits} 
needs complicated formulation
since the superstring field theory suffers
from many problems as picture-changing interactions
and Lorentz invariance.\cite{Mandelstam,Green,Berkovits}

On the other hand, the non-perturbative analysis for the string theory
has been developed recently.
The new ideas as D-branes and M-theory give
more profound understanding for the strongly coupled string theory.
Therefore, there are many approach for the description of M-theory.
Some of them are Matrix theory\cite{BFSS}
and Matrix string theory.\cite{MOTL,BANKS-SEIBERG,DVV}
Especially, Matrix string theory describes M-theory
in terms of a $1+1$ dimensional super Yang-Mills theory and
this theory goes to the conformal field theory on the symmetric product
space in the IR limit.
The IR dynamics of the matrix string was well-analyzed 
in terms of super Yang-Mills theory.\cite{Vanhove,sugino0}
There are many progresses in this direction 
recently.\cite{Smolin,Sugino,Sekino-Yoneya}

The analysis of
the string theory on the symmetric product space was given for
the closed string case.\cite{DMVV}\cite{Bantay}
The twisted sector can be interpreted physically as the {\it long string}
which is constituted from the {\it string bits} by connecting their edges.
This interpretation can be established further by evaluating the
partition function.
The generating function of the partition function is expressed by
exponential lifting of the one-body partition function.
This nature of the partition function is the string theory version of
the sum of the connected Feynman diagrams for vacuum
in the quantum field theory.
As discussed by L.Susskind\cite{Susskind},
this partition function has the
structure of the discrete lightcone quantization (DLCQ).
Thus this theory can be interpreted as the second-quantized
string theory in DLCQ.

In the previous work\cite{Matsuo-Fuji}, we developed these ideas into the
bosonic open string theory.
In the open string case, the twisted sector on the orbifold singularity
of the symmetric product can be also interpreted as the long strings.
But these long strings have two types, the long open strings and the long
closed strings.
Furthermore, the annulus partition function can be interpreted as
the annulus, M\"obius strip, Klein bottle and torus partition function
in terms of the long string sense.
Therefore this theory can be interpreted
as the second quantization of the unoriented open/closed string theory.
To establish this structure, we constructed the boundary states
for this theory.
These boundary states are classified into three types,
the boundary states, the cross-cap states, and the joint states.
Here the joint states describes the connection of two strings at the
boundary.
The appearance of the cross-cap states means that this theory describes
the unoriented theory and joint state implies the possibility of
open/closed interactions.
So the physical interpretation can be explained naturally from the
closed string sector.
Finally we discussed the consistent Chan-Paton gauge group
for the anomaly-free theory.
This group can be determined from the dilaton tadpole cancellation,
and the result was $SO(2^{13})$ as ordinary string theory.

Generally speaking, the representation of
the second-quantized string theory by the symmetric product is
much simpler than the string field theory.
Especially, as the superstring field theory has complicated formulation,
we aim to describe open superstring theory in simple way.
In this paper, we extended the ideas in the previous work\cite{Matsuo-Fuji}
to the open superstring case.
We found the boundary condition for fermionic fields
on the non-abelian orbifold by the $\bZ_2$  extension of the
bosonic boundary conditions and classified the solutions for
the consistency conditions of the permutation orbifold.
For these classified boundary conditions. we calculated
the partition function and examined that
the annulus, M\"obius strip, Klein bottle and torus partition function
for the superstring appears.
In the closed string sector, we constructed the boundary/cross-cap states
for the fermionic fields.
The fermionic boundary states also describe the boundary, cross-cap
and joint states and explain the change of the topology of the world-sheet.
Finally, from the dilaton tadpole cancellation condition,
the Chan-Paton group can be determined to be $SO(32)$ as ordinary
Type I superstring.

The organization of this paper is as follows.
In section 2, we review the closed string on the symmetric product
space.
Here we calculate the partition function for the superstring in detail
as the preparation for the open string calculation.
In section 3, we consider the consisteny condition
for the periodicities and boundary conditions of non-abelian orbifold
and classify the solutions.
Then we calculate the fermionic partition function for the irreducible
sets.
These considerations are made for the annulus, M\"obius strip and Klein
bottle diagrams.
These results are summarized in the final subsection and we will
calculate the the generating function for the partition functions.
In section 4, we construct the boundary/cross-cap states for fermionic
string.
After the classification for the irreducible combination of the
periodicity and the boundary condition,
we calculate the inner products of the boundary states which
correspond to the situation of section 3.
In section 5, we consider the dilaton tadpole cancellation condition
and determine the consistent Chan-Paton gauge group.
In section 6, we discuss about the interaction briefly and
show some future direction of this work.

\section{Closed Superstring on Symmetric Product}

The closed string theory on the symmetric product space
describes the second-quantized closed string theory.\cite{DMVV}
In this section we will review this nature and calculate the partition
function in detail for the prepartion of the open string case.

\subsection{Hilbert space of closed string on symmetric product}

Let $M$ be some manifold. The symmetric product of this space is
defined as $S^NM\equiv M^{\otimes N}/S_N$, where $S_N$ denotes
the permutation group.
On this symmetric product space, the twisted sector of the
the bosonic field $\bX^{I}(\sigma)$ ($I=0,\cdots,N-1$) which
belongs to the Hilbert space $\cH_N$ is defined as,
\begin{equation}
\bX^{I}(\sigma+2\pi)=(h\cdot\bX)^{I}(\sigma), \quad
h\in S_N,
\end{equation}
where we denoted as, $h\cdot \bX\equiv \sum_Ih^{I,J}X^J$ and
$\sigma$ represents the space-like coordinate of the world-sheet.
This Hilbert space can be decomposed into that of the twisted sectors
of the conjugacy classes.
The conjugacy class $[h]$ for the permutation group $S_N$ is
classified by the partition of $N$ and written as the product of the
elements of the cyclic permutation group $(n)$ as,
\begin{eqnarray}
[h] &\equiv& (1)^{N_1}(2)^{N_2}\cdots(k)^{N_k},
\label{conjugacy} \\
N&=&\sum_{n=1}^{k}nN_n.
\end{eqnarray}
Thus the Hilbert space is decomposed as,
\begin{equation}
\cH_N=\bigoplus_{\scriptsize \begin{array}{c}
\{N_n\}\\N=\sum_{n}nN_n\end{array}}\cH_{\{N_n\}},
\end{equation}
where the Hilbert space $\cH_{\{N_n\}}$ represents
the twisted sector of the conjugacy class.

In general, the orbifold partition function which is invariant
under the group action $\Gamma$ can be obtained by summing over all twisted
sectors as,
\begin{eqnarray}
Z(\tau,\bar{\tau})&=&
\frac{1}{|\Gamma|}\sum_{h,g\in \Gamma}\chi_{h,g}, \\
\chi_{h,g}(\tau,\bar{\tau})&\equiv&
{\rm Tr}_{\cH_h}g\be{\tau L_0-\tilde{\tau}\tilde{L}_0},
\\
\be{x}&\equiv&e^{2\pi ix}.
\end{eqnarray}
From the consistency of the path integral,
in the case of the non-abelian orbifold, we need to restrict above
summation to the elements which satisfied the condition,
\begin{equation}
hg=gh.
\end{equation}
Therefore, we have to choose the twist along the time direction
as the centralizer group $\cC_h$ for the the conjugacy class $[h]$.
The centralizer group can be written as,
\begin{equation}
g\in S_{N_1}\times (S_{N_2}\times \bZ_{2})\cdots (S_{N_k}\times \bZ_{k}).
\label{centralizer}
\end{equation}
Thus the partition function for the symmetric product space is written
as,
\begin{eqnarray}
Z_N(\tau,\bar{\tau})&=&
\frac{1}{|S_N|}\sum_{\scriptsize \begin{array}{c}
g,h\in S_N\\hg=gh
\end{array}}\chi_{h,g}(\tau,\bar{\tau}) \nn \\
&=&
\sum_i\frac{1}{|\cN_i|}\sum_{g\in \cN_i}\chi_{h,g}(\tau,\bar{\tau}),
\end{eqnarray}
where $h$ belongs to the conjugacy class $\cC_i$ and
we denoted $\cN_i$ as its centralizer group and summation is taken over
all conjugacy class.
In the derivation of above representation, we used the relation
$|S_N|=|\cC_i|\cdot|\cN_i|$.
The number of the elements of the contralizer group
(\ref{centralizer}) is,
\begin{equation}
|\cN_h|=\prod_{n=1}^{k}n^{N_n}N_n!.
\end{equation}

The physical interpretation of this twisted sector can be given
as follows.
Let $h$ be the cyclic permutation of $n$ elements $\cycl_n$,
the periodicity for the bosonic field $\bX^I$ ($I=0,\cdots,n-1$)
along the space direction is,
\begin{equation}
\bX^{I}(\sigma+2\pi)=\bX^{I+1}(\sigma),
\end{equation}
where $I$ is defined mod $n$.
This means that $n$ short strings are connected to form the long stirng
of length $n$,
\begin{equation}
\bX^{I}(\sigma+2\pi n)=\bX^{I}(\sigma).
\label{long period}
\end{equation}
These short strings are sometimes called ``{\it srting bits}''.
For the general element (\ref{conjugacy}),
$N$ short strings form $N_n$ long strings of length $n$.

For the fermionic fields $\Psi$ on the permutation orbifolds
the twisted sector which corresponds to above bosonic fields
can be expressed as,
\begin{equation}
\Psi^{I}(\sigma+2\pi)=\be{\frac{\alpha_{I}}{2}}\Psi^{I+1}(\sigma).
\end{equation}
$\alpha_I$ is $0$ or $1$.
The physical meaning for this twisted sector is that
$n$ short NS or R string are connected to form the
long fermionic string of length $n$ which has the peridicity as,
\begin{eqnarray}
\Psi^{I}(\sigma+2\pi n)&=&\be{\frac{\alpha}{2}}\Psi^{I}(\sigma),\\
\alpha&\equiv&\sum_{I=0}^{n-1}\alpha_I, \quad \mod{2}.
\end{eqnarray}

The action of the centralizer group (\ref{centralizer}) is has clear
interpretation.
The factor $\bZ_n$ means the rotation of the string bits that constitute
a long string of length $n$.
The factor $S_{N_n}$ reshuffle the long strings of the same length $n$.

 \begin{figure}[ht]
  \centerline{\epsfxsize=7cm \epsfbox{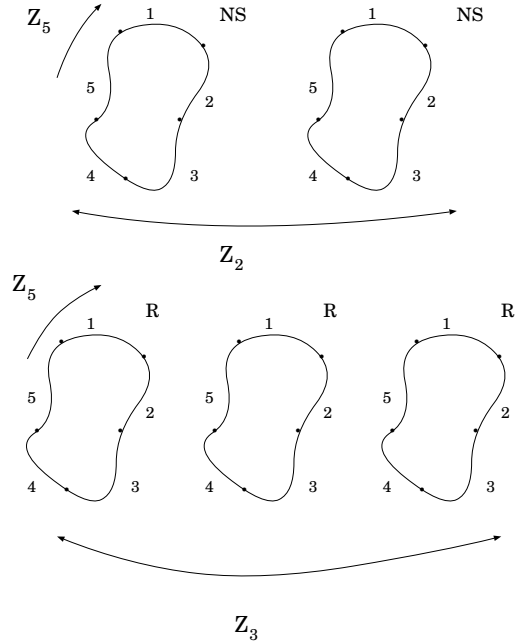}}
  \vskip 3mm
  \caption{Same type of long strings are reshuffled in each irriduceble sets}
 \end{figure}

In order to evaluate the partition function explicitly,
we should decompose these elements into the ``irreducible sets''.
In the element of the centralizer group, $S_{N_n}$ can be
decomposed further into its conjugacy class as,
\begin{eqnarray}
&&[S_{N_n}]=(1)^{M_{n,1}}(2)^{M_{n,2}}\cdots ,\\
&&\sum_{s_n>0}s_nM_{n,s_n}=N_n.
\end{eqnarray}
The ``irreducible set'' is the pair $h\in \bZ_n^{\otimes s_n}$,
$g\in \bZ_n^{\otimes s_n}\times \bZ_{s_n}$.

For the fermionic field, the action of the centralizer group is
more complicated than the bosonic case.
From the consistency condition $hg=gh$,
the action of the centralizer group is the rotations and reshuffling of
the long strings which have the same length and belong to the same
fermionic sector, and appropriate $\bZ_2$ action.
In the next subsection, we will discuss
the action of the centralizer group in more detail.

Thus we see how the conjugacy classes and its centralizers decompose
into the irreducible sets.
The full partition function is evaluated as the sum of the product of
the partition function for the irreducible sets.

\subsection{Partition function for closed string}
\subsubsection{Fermionic partition function for the irreducible set}

To consider the partition ffunction of the superstring theory on
$S^N\bR^8$, we have to evaluate that of the complex fermionic fields.
For the fermionic fields,
the orbifold group is extended to
$S_N\times \bZ_2^{\otimes N}$.

The irreducible sets can be reduced to $nm$ complex fermionic free fields
where $h$ and $g$ act as,
\begin{eqnarray}
 h & = & {\rm diag}(A_0\cycl_n,\cdots,A_{m-1}\cycl_n),\nn\\
 g & = & {\rm diag}(B_0\cycl_n^{p_0},\cdots,B_{m-1}\cycl_n^{p_{m-1}})\cdot
 \cyclb_m\,\,,
\end{eqnarray}
where $\cycl_n$ and $\cyclb_m$ acts on $nm$ free fermionic fields
$\Psi^{I,J}$ ($I=0,\cdots ,n-1$ and $J=0,\cdots ,m-1$) as,
\begin{eqnarray}
(\cycl_n \cdot \Psi)^{I,J} &=& \Psi^{I+1,J}, \nn\\
(\cyclb_m \cdot \Psi)^{I,J} &=& \Psi^{I,J+1}.
\end{eqnarray}
$A_{J}$'s and $B_{J}$'s are $n\times n$ matrices
$A_J=\diag(\be{\frac{\alpha_{0,J}}{2}},\cdots
,\be{\frac{\alpha_{n-1,J}}{2}})$,
\\
$B_J=\diag(\be{\frac{\beta_{0,J}}{2}},\cdots
,\be{\frac{\beta_{n-1,J}}{2}})$
where $\alpha_{I,J}$ and $\beta_{I,J}$ takes their values in $0$ or $1$.
The actions of $h$ and $g$ on $\Psi$ are written asas,
\begin{eqnarray}
 (h\cdot \Psi)^{I,J} &=& \be{\frac{\alpha_{I,J}}{2}}\Psi^{I+1,J}, \nn\\
 (g\cdot \Psi)^{I,J} &=& \be{\frac{\beta_{I,J}}{2}}\Psi^{I+p_J,J+1},
\end{eqnarray}
The consistency condition $hg=gh$ becomes nontrivial conditions
for $\alpha$'s and $\beta$'s as,
\begin{equation}
\beta_{I,J} + \alpha_{I+P_J,J+1} = \alpha_{I,J} +\beta_{I+1,J}.
\label{torus consistency}
\end{equation}
Taking a sum of $I$ in the above relation,
the condition for $\alpha_J \equiv \sum_{I=0}^{n-1}\alpha_{I,J}$ is
\begin{equation}
\alpha_J =\alpha_{J+1}.
\end{equation}
Thus $\alpha_J$'s do not depend on $J$.
This fact means that
all long strings in the irreducible sets
have same periodicity on space direction.
So we define $\alpha \equiv \alpha_J$
where $\alpha$ is defined modulo 2.

By diagonalizing the action for $h$,
we should introduce the basis for the fermionic field $\psi^{a,J}$
which is constructed from the discrete Fourier transformation of
the fermionic field $\Psi^{I,J}$ as,
\begin{equation}
\psi^{a,J}\equiv\frac{1}{\sqrt{n}}\sum_{I=0}^{n-1}\be{-\frac{aI}{n}
-\frac{\alpha I}{2n}+\sum_{K=0}^{I-1}\frac{\alpha_{K,J}}{2}}\Psi^{I,J},
\quad a=0,\cdots n-1.
\end{equation}
The mode expansion for $\psi^{a,J}$ is
\begin{equation}
\psi^{a,J}=i^{-1/2}\sum_{r\in \bZ}\left(
\psi_{r+\frac{a}{n}+\frac{\alpha}{2n}}^{a,J}
e^{i(r+\frac{a}{n}+\frac{\alpha}{2n})w}
+\tilde{\psi}_{r-\frac{a}{n}-\frac{\alpha}{2n}}^{a,J}
e^{-i(r-\frac{a}{n}-\frac{\alpha}{2n})\bar{w}}
\right),
\label{mode T2}
\end{equation}
where $w=\sigma^{1}+i\sigma^{0}$.
Since the fermionic fields are complex fields,
there are also conjugate fields $\bar{\Psi}^{I,J}$.
The mode expansion for the conjugate field $\bar{\psi}^{a,J}$ is
\begin{equation}
\bar{\psi}^{a,J}=i^{-1/2}\sum_{r\in \bZ}\left(
\bar{\psi}_{r-\frac{a}{n}-\frac{\alpha}{2n}}^{a,J}
e^{i(r-\frac{a}{n}-\frac{\alpha}{2n})w}
+\tilde{\bar{\psi}}_{r+\frac{a}{n}+\frac{\alpha}{2n}}^{a,J}
e^{-i(r+\frac{a}{n}+\frac{\alpha}{2n})\bar{w}}
\right).
\end{equation}
The commutation relations are
\begin{equation}
\{\psi_{r+\frac{a}{n}+\frac{\alpha}{2n}}^{a,J},
\bar{\psi}_{s-\frac{b}{n}-\frac{\alpha}{2n}}^{b,L}\}
=\delta_{J,L}\delta_{r+s,0}\delta_{a,b}.
\end{equation}

The action of $g$ on this diagonalized basis $\psi^{a,J}$ becomes,
\begin{equation}
(g\cdot \psi)^{a,J}
=
\be{\frac{\beta_{0,J}}{2}
-\sum_{K=0}^{p_J-1}\frac{\alpha_{K,J+1}}{2}
+\frac{ap_J}{n}+\frac{\alpha p_J}{2n}} \psi^{a,J+1} .
\end{equation}
The eigenvalues for $g$ action are evaluated as,
\begin{eqnarray}
&&\mu^{a,b}=\be{\frac{\beta^{\{p\}}}{2m}+\frac{ap}{nm}+\frac{\alpha p}{2nm}
+\frac{b}{m}}, \\
&& p\equiv \sum_{J=0}^{m-1}p_J , \quad
\beta^{\{p\}}\equiv
\sum_{J=0}^{m-1}\left(\beta_{0,J}-\sum_{K=0}^{p_J-1}\alpha_{K,J+1}\right),
\\
&&a=0,\cdots ,n-1 , \quad b=0,\cdots ,m-1. \nn
\end{eqnarray}
The action of $g$ on diagonalized oscillator
$\psi^{a,b}_{r+a/n+\alpha /2n}$ is written as,
\begin{equation}
(g\cdot \psi)^{a,b}_{r+a/n+\alpha /2n}=\mu^{a,b}\psi^{a,b}_{r+a/n+\alpha
/2n}.
\end{equation}

Using these diagonalized fields,
the (chiral) oscillator part of the partition function
can be evaluated as,
\begin{eqnarray}
&&\prod_{a=0}^{n-1}\prod_{b=0}^{m-1}\prod_{r=1}^{\infty}\left(
1-\be{\frac{\beta^{\{p\}}}{2m}+\frac{ap}{nm}+\frac{\alpha p}{2nm}
+\frac{b}{m}} \be{(r-a/n-\alpha /2n)\tau}
\right), \nn \\
&=&
\prod_{a=0}^{n-1}\prod_{r=1}^{\infty}\left(
1-\be{\frac{\beta^{\{p\}}}{2}+\frac{ap}{n}+\frac{ap}{2n}}
\be{(r-a/n-\alpha /2n)m\tau }
\right), \nn \\
&=&
\prod_{a=0}^{n-1}\prod_{r=1}^{\infty}\left(
1-\be{\frac{\beta^{\{p\}}}{2}}\be{-p(r-a/n-\alpha /2n)}
\be{(r-a/n-\alpha /2n)m\tau }
\right), \nn \\
&=&
\prod_{s\in \bN -\alpha/2}\left(
1-\be{\frac{\beta^{\{p\}}}{2}}\be{\Bigl(\frac{m\tau -p}{n}\Bigr)s}
\right).
\end{eqnarray}
Thus the partition function for the $nm$ fermionic strings in the
irreducible set are knitted together to give the partition function
for the one long fermionic string on the torus
with modified moduli parameter $\tau_{n,m,p}$,
\begin{equation}
\tau_{n,m,p} \equiv \frac{m\tau -p}{n}.
\end{equation}

To evaluate the full partition function for the irreducible set,
we need to add some important factors.
At first, we consider the multiplicity of $\alpha_{J}$'s and $\beta_{J}$'s.
If we sum over all $\alpha_{J}$'s and $\beta_{J}$'s,
the value of $\alpha$ and $\beta^{\{p\}}$ can be taken freely.
So under fixing $\alpha$ and $\beta^{\{p\}}$, the number of
the multiplicity of $\alpha_{J}$'s and $\beta_{J}$'s is $2^{nm-1}$.
Thus we have to add the factor as
($Z_2$ is the action of fermionic boundary condition),
\begin{equation}
\frac{2^{nm-1}}{|Z_2|^{nm}} = \frac{1}{2}.
\end{equation}
And we replace $\beta^{\{p\}}$ to $\beta$.

Next we consider the zero-point energy.
The zero-point energy for the fermionic field
which has the fractional moding as $\bZ-\theta$ is\cite{Polchinski},
\begin{equation}
-\frac{1}{2}\omega^{\theta}\equiv -\frac{1}{2}\sum_{n=1}^{\infty}(n-\theta)
=-\frac{1}{48}-\frac{1}{16}(2\theta-1)^2.
\end{equation}
Therefore
the zero-point energy of a (complex) fermionic string with length $n$
on $S^N\bf{R}^2$ is
\begin{eqnarray}
&&
\sum_{a=0}^{n-1}\left( 
-\frac{1}{2}\omega^{\theta =\frac{a}{n}+\frac{\alpha}{2n}}
-\frac{1}{2}\omega^{\theta =1-\frac{a}{n}-\frac{\alpha}{2n}}
\right), \nn \\
&=&
\sum_{a=0}^{n-1}\left(
-\frac{1}{24} + \frac{1}{8}\Bigl\{2(\frac{a}{n}+\frac{\alpha}{2n})-1
\Bigr\}^2
\right), \nn \\
&=& \frac{1}{12n}-\frac{\alpha}{4n}+\frac{\alpha ^2}{8n}.
\end{eqnarray}
Thus the zero-point energy for the long string with length $n$ 
is $\frac{1}{n}$ of short string zero-point energy.

Here we consider the oscillator part of 
conjugate fermionic fields $\bar{\Psi}^{I,J}$.
The calculation is the same as that of $\Psi^{I,J}$ .
The result is 
\begin{equation}
\prod_{s\in \bN -(1-\alpha/2)}\left(
1-\be{\frac{-\beta^{\{p\}}}{2}}\be{\Bigl(\frac{m\tau -p}{n}\Bigr)s}
\right).
\end{equation}

Finally we consider the various factors that comes from the 
vacuum.
Since the vacuum for the $\psi^{a,b}_{r-a/n-\alpha /2n}$
($0\le a/n+\alpha /2n \le 1/2$) has charge
$\frac{1}{2}-\frac{a+\alpha /2}{n}$,
the contribution from the ground state charge for
the irreducible set is evaluated as,
\begin{equation}
{\bf e}\Bigl[
\frac{(\beta^{\{p\}}-1)(1-\alpha)}{8}
-\frac{p}{8n}\left(
\alpha ^2-2\alpha
\right)
-\frac{(\tilde{\beta}^{\{p\}}-1)(1-\tilde{\alpha})}{8}
+\frac{p}{8n}\left(
\tilde{\alpha} ^2-2\tilde{\alpha}
\right)
\Bigr].
\end{equation}
For the irreducible set of $nm$ strings, $m$ long string
with length $n$ moving coherently.
Its toggled vacuum is assigned to the toggled world-sheet
as ordinary strings.
So the fermion number of the vacuum and the space-time fermion number
for the various sector is the same as that of
ordinary string theory.

Gathering all these factors,
the full (chiral) fermionic partition function for the irreducible set
of $S^{N}{\bf R}^8$ is expressed as follows.\cite{Polchinski,Lang}

{\it Type IIA,B}
\begin{eqnarray}
&&
Z^{II}(n,m,\{p\}|\tau,\bar{\tau})=
Z^{+}(\tau_{n,m,p})Z^{\pm}(\bar{\tau}_{n,m,p})^{*}, \\
&&
Z^{\pm}(\tau)=\frac{1}{2}[Z^{0}_{0}(\tau)^4-Z^{0}_{1}(\tau)^4
-Z^{1}_{0}(\tau)^4
\mp Z^{1}_{1}(\tau)^4],
\label{super partition fcn}\\
&&
Z^{a}_{b}(\tau)\equiv q^{(3a^2-1)/24}
\be{ab/8}\nn \\
&& \quad\quad\quad\quad \times
\prod_{m=1}^{\infty}\Bigl(1-\be{b/2}q^{m-(1-a)/2}\Bigr)
\Bigl(1-\be{-b/2}q^{m-(1+a)/2}\Bigr),
\end{eqnarray}
where $q\equiv \be{\tau}$.

{\it Type 0A,B}
\begin{equation}
Z^{0}(n,m,\{p\}|\tau,\bar{\tau})=
\frac{1}{2}[|Z^{0}_{0}|^N+|Z^{0}_{1}|^N+|Z^{1}_{0}|^N+|Z^{1}_{1}|^N].
\end{equation}
where $N=8$.

Thus we obtained the partition function for the irreducible set
of the $S^{N}\bR^8$.

\subsubsection{Generating function of the partition function}

As we obtained the partition function for the irreducible sets,
we evaluate the partition function on $S^N\bR^8$
by summing all products of irreducible sets.
The combinatorial feature of the partition function for the irreducible
set is same as that of bosonic case.\cite{Matsuo-Fuji}

The partition function for each conjugacy class (\ref{conjugacy})
can be written as
follows,
\begin{eqnarray}
\chi_{h,g} &=& \prod_{n,s_n>0}(n^{s_n}Z(n,s_n))^{M_n,s_n}, \\
Z(n,m) &=& \frac{1}{n^{m}}\sum_{\{p\}}Z(n,m,\{p\}|\tau,\bar{\tau})
=\frac{1}{n}\sum_p Z(n,m,p|\tau,\bar{\tau}).
\end{eqnarray}
Gathering all factors,
the whole partition function is written as,
\begin{eqnarray}
Z_{N}(\tau,\bar{\tau}) &=&
\sum_{\scriptsize
\begin{array}{c}
N_n,M_{n,m}\geq 1\\
\sum n N_n=N\\
\sum m M_{n,m}=N_n
\end{array}}
\prod_{n,m \geq 1}\frac{1}{M_{n,m}!}
\left(
\frac{1}{m} Z(n,m|\tau,\bar{\tau})
\right)^{M_{n,m}}.
\end{eqnarray}

The generating function of the partition function  can be
evaluated as,
\begin{eqnarray}
Z(\zeta|\tau,\bar{\tau})&\equiv&
\sum_{N=0}^{\infty}\zeta^{N}Z_{N}(\tau,\bar{\tau}), \nn \\
& = & \exp\left(
\sum_{N=1}^\infty \zeta^N
\cV_N\cdot Z_1(\tau,\bar\tau)
\right),\\
\cV_N\cdot f(\tau,\bar\tau) &\equiv &
\frac{1}{N} \sum_{\scriptsize \begin{array}{c}
 a,d=1,\cdots,N\\b=0,\cdots,d-1\\ad=N
  \end{array}}
f\left(\frac{a\tau+b}{d},\frac{a\bar\tau+b}{d}\right)
\label{e_Hecke}
\end{eqnarray}
The operator $\cV_N$ is called Hecke operator\cite{SERRE,apostol}
which maps a modular form to another one with the same weight.

In large $N$ limit, this partition function becomes as,
\begin{equation}
\sum_{m=1}^{\infty}\frac{1}{nm}\sum_{p=0}^{n-1}Z_1(\tau,\bar{\tau})
\to
\int\frac{d^2\tau}{{\rm Im}\tau}Z_1(\tau,\bar{\tau}).
\end{equation}
Thus our partition function can be interpreted as the discretized
version of the ordinary amplitude for the light-cone quantization.
As we reproduced the integral of moduli in the continuum limit,
we set the redundant variable $\tau$ to $\tau=i$.

\section{Open Superstring on Symmetric Product}
We will consider
the open superstring theory on the symmetric product space.
The nature of the second-quantized open string can be established by
the evaluation of the partition function.
Since the Hilbert space of the open string sector can be determined
by specifying the boundary conditions, we have to calssify
the boundary condition.
The twisted sector in the open string theory on the abelian orbifold
is well-known.\cite{Pradisi-Sagnotti,Ishibashi Ph.D}
Although the boundary condition for the abelian orbifold reduces to
Neumann or Dirichlet condition,
the boundary condition for the non-abelian orbifold has
various sectors.
The consistency condition for the bosonic boundary condition
on the non-abelian orbifold is discussed 
in the previous work.\cite{Matsuo-Fuji}
In this article, we generalize the previous results to
fermionic case and then calculate the partition function for
the permutation orbifold.

\subsection{Open string Hilbert space}

The Hilbert space for the open string is specified by the
two boundary conditions.
In this subsection, we concentrate on the fermionic fields.
The fermionic action for the ordinary string is,
\begin{equation}
S=-\frac{i}{2\pi}\int_{0}^{\pi}d\sigma^1\int_{-\infty}^{\infty}d\sigma^0
\Psi(\sigma^0,\sigma^1)
\rho_{\alpha}\partial_{\alpha}\Psi(\sigma^0,\sigma^1),
\end{equation}
where $\rho_{\alpha}$ is two-dimensional Dirac matrix.
The translation invariance of the action gives the boundary condition
for the fermionic field,
\begin{equation}
\Psi_L(\sigma^0,\sigma^1_0)\pm f\Psi_R(\sigma^0,\sigma^1_0)=0,
\quad f\in \bZ_2,
\end{equation}
where $\sigma^1_0$ is the boundary value of $\sigma^1$.
The sign $\pm$ in the boundary condition are referred as
Neumann and Dirichlet condition.
When the values of $f$ take different value at two boundary,
this sector is called Neveu-Schwarz (NS) sector and
when the values of $f$ take same value at two boundary,
this sector is called Ramond (R) sector.
Thus the Hilbert space of the open string is specified by the
boundary conditions.

Next we will discuss the case that the target space is the orbifold
$M/\Gamma$.
At the fixed point of the action of the discrete group $\Gamma$,
the boundary condition for the fermionic field is modified as,\cite{Harvey-Minahan}
\begin{equation}
\Psi_{L}(\sigma^0,\sigma^1_0)\pm
f\cdot \Psi_R(\sigma^0,\sigma^1_0)=0,  \quad
f \in \Gamma\times \bZ_2. \label{fermionic bdy cond}
\end{equation}
We assume that the left- and right-mover have the same form of the
boundary condition for the orbifold action and
admit the asymmetry for the fermionic $\bZ_2$ part,
\footnote{
Here we imposed the $\bZ_2$ extended constraint for the fermionic
boundary condition.
In the permutation orbifold case,
if one want to realize only the long closed string
which has the same periodicity,
one can impose the completely left-right symmetric constraint.
}
\begin{equation}
f^2=\cE, \quad \cE\in\Gamma\times \bZ_2,
\label{LR symmetry F}
\end{equation}
where $\cE$ belongs to the identity element in $\Gamma$.

For the permutation orbifold, the boundary twist $f$ belongs to the
conjugacy class as,
\begin{equation}
f\in \left[(1)^{N_1}(2)^{N_2}\right]\times\bZ_2^{\otimes N},
\end{equation}
where $N=N_1+2N_2$.
The physical interpretation of this boundary twist is as follows.
The boundary twist in the element of $(1)$ expresses the loose end of the
string.
In this article, we assign Neumann condition
for this boundary condition to discuss the fundamental string.
On the other hand,
the boundary twist if the element of $(2)$ expresses the connection of two
strings at this
boundary.
In this article, we assign Dirichlet condition
for this boundary condition because two strings are connected smoothly.

 \begin{figure}[ht]
  \centerline{\epsfxsize=10cm \epsfbox{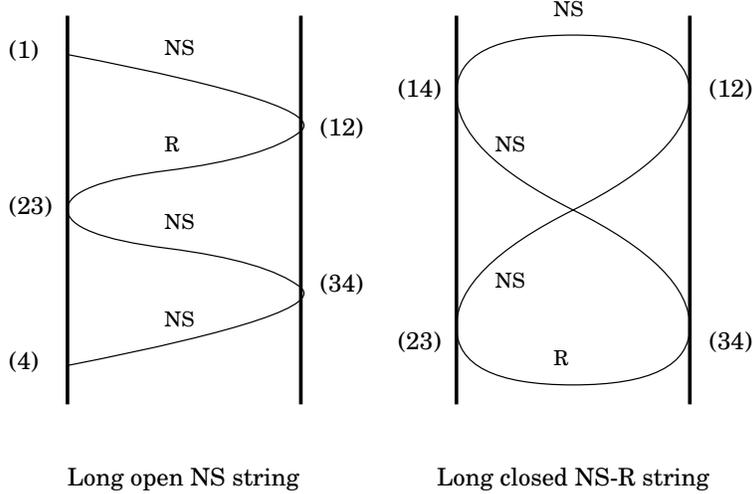}}
  \vskip 3mm
  \caption{Long open (closed) superstrings}
 \end{figure}

The open string Hilbert space for the permutation orbifold
can be specified by choosing the two boundary twists $f_1$, $f_2$
as above.
For example, we illustrate the situation
for a long open string (Figure 2 left),
\begin{eqnarray}
f_1 &=& \left[+\cycl_1\right]\otimes
\left[{\rm diag}(-1,-1) \cycl_2\right]
\otimes \left[-\cycl_1\right],
\nn \\
f_2 &=& \left[{\rm diag (-1,-1)}\cycl_2\right]
\otimes \left[{\rm diag}(+1,+1) \cycl_2 \right].
\nn
\end{eqnarray}
Three NS open string bits and one R open string bit are connected
and constitute a NS long open string.
The situation for a long closed string is (Figure 2 right),
\begin{eqnarray}
f_1 &=&  \left[{\rm diag}(+1,+1) \cycl_2 \right]
\otimes \left[{\rm diag}(-1,-1)\cycl_2 \right],
\nn \\
f_2 &=& \left[{\rm diag (-1,-1)}\cycl_2 \right]
\otimes \left[{\rm diag}(-1,+1) \cycl_2 \right].
\nn
\end{eqnarray}
Three NS open string bits and one R open string bit are connected
and constitute a NS-$\widetilde{{\rm R}}$ long closed string.
It is clear that we may realize the long open and closed string
of the arbirary length by the pair of the boundary twists $(f_1,f_2)$.

\subsection{Partition function for Annulus diagram}

The partition function for the oriented open superstring
is defined on the  annulus diagram,
\begin{equation}
{\rm Tr}_{\cH_{f_1,f_2}}(g\be{\tau L_0}).
\end{equation}
Here and in the following sections, the moduli parameter $\tau$ 
for the annulus is pure imaginary.
We will discuss the partition function of the fermionic field
in this section to obtain the superstring partition function.

The boundary conditions for the fermionic field are written as,
\begin{eqnarray}
\Psi_L(\sigma^0,0)&=&f_1\cdot\Psi_R(\sigma^0,0),\nn \\
\Psi_L(\sigma^0,\pi)&=&f_2\cdot\Psi_R(\sigma^0,\pi).
\label{boundary cond F}
\end{eqnarray}
To write the mode expansion for this field,
a chiral field $\Psi$ should be introduced on the double cover of the
annulus diagram (Figure 3).\cite{Pradisi-Sagnotti}

 \begin{figure}[ht]
  \centerline{\epsfxsize=6cm \epsfbox{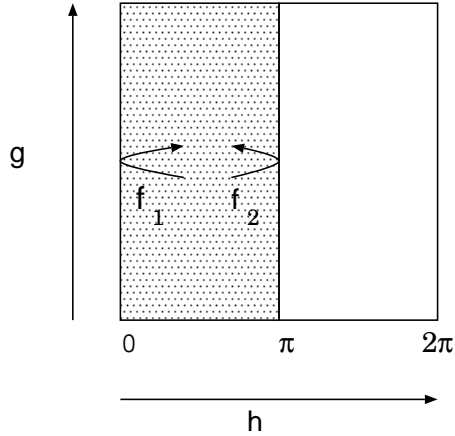}}
  \vskip 3mm
  \caption{Double cover for annulus diagram}
 \end{figure}

This chiral field has following periodicities,
\begin{equation}
\Psi(\sigma^0,\sigma^1+2\pi)=h\cdot \Psi(\sigma^0,\sigma^1), \quad
\Psi(\sigma^0+2\pi,\sigma^1)=g\cdot \Psi(\sigma^0,\sigma^1).
\end{equation}
For the consistency of the path integral, $h$ and $g$ must satisfy
\begin{equation}
hg=gh.
\label{hg cond A}
\end{equation}
From the boundary condition at $\sigma^1=\pi$,
the left- and right-moving fields are identified to this chiral field as,
\begin{equation}
\Psi_L(\sigma^0,\sigma^1)=\Psi(\sigma^0,\sigma^1), \quad
\Psi_R(\sigma^0,\sigma^1)=f_2\cdot \Psi(\sigma^0,2\pi-\sigma^1),
\quad 0\le \sigma^1 \le \pi.
\end{equation}
To satisfy the boundary condition at $\sigma^1=0$,
$h$ and $f$'s must be related as,
\begin{equation}
 h=f_2^{-1}\cdot f_1.
\label{h cond A}
\end{equation}

The consistency condition for the periodicity along the time direction and
the boundary conditions is given as,
\footnote{
We note that this condition is less restrictive than the bosonic one.
If we consider Type 0 theory, we can use the bosonic condition
(\ref{fg cond A}).
But in order to realize the NS-$\widetilde{{\rm R}}$
and R-$\widetilde{{\rm NS}}$ sector, we have to impose this condition.
In the construction of the boundary state,
we can understand the reason why this codition should be imposed.
}
\begin{equation}
g\cdot f_1 = \cD_1 \cdot (f_1 \cdot g), \quad
g\cdot f_2 = \cD_2 \cdot (f_2 \cdot g), \label{fg cond A}
\end{equation}
where $\cD_1$, $\cD_2$ are $\bZ_2$ twists.

In the case of permutation group $\Gamma=S_N$,
there are irreducible set of the twists $h,g,f_1,f_2$.
The irreducible set of the solution for  the consistency conditions
are classified into three types in bosonic case.\cite{Matsuo-Fuji}
\begin{itemize}
\item $I_A$ type solution:
\begin{eqnarray}
h&=&\diag(\overbrace{\cycl_n,\cdots,\cycl_n}^{m}),
\nn \\
g&=&\diag(\cycl_n^{p_0},\cdots,\cycl_n^{p_{m-1}})\cdot\cyclb_m,
\nn \\
f_1&=&\diag(\cycl_n^{q_0}\inv_n,\cdots,\cycl_n^{q_{m-1}}\inv_n),
\nn \\
f_2&=&\diag(\cycl_n^{q_0+1}\inv_n,\cdots,\cycl_n^{q_{m-1}+1}\inv_n),
\end{eqnarray}
where $p_J,q_J$ take their values in $0,1,\cdots,n-1$ (mod $n$) and 
$q_J$'s satisfy the condition, 
\begin{equation}
q_{J+1}-q_J\equiv 2p_J \mod n.
\label{pq cond IAB}
\end{equation}

\item $II_A$ type solution:

For even $m$, there is a solution which has the same form of $h$ and $g$ as
$I_A$ case and $f_i$ takes the form as,
\begin{eqnarray}
f_1&=&\diag(\cycl_n^{q_0}\inv_n,\cdots,\cycl_n^{q_{m-1}}\inv_n)
\cdot(\cyclb_{m/2})^{m/2}, \nn \\
f_2&=&\diag(\cycl_n^{q_0+1}\inv_n,\cdots,\cycl_n^{q_{m-1}+1}\inv_n)
\cdot(\cyclb_{m/2})^{m/2}, 
\end{eqnarray}
where $p_J$ and $q_J$ satisfy the conditions,
\begin{eqnarray}
&&
q_{J+1}-q_J\equiv p_J+p_{J+m/2}, \nn \\
&&
q_{J+m/2}\equiv q_{J}, \quad \mod n.
\label{pq cond IIAB}
\end{eqnarray}

\item $\widetilde{II}_A$ type solution:

For even $m$, there is another solution which has the same form of 
$h$ and $f_1$,$f_2$ as $II_A$ case and $g$ takes the form as,
\begin{equation}
g=\diag(\cycl_n^{p_0},\cdots,\cycl_n^{p_{m-1}})\cdot
\diag(\cyclb_{m/2},\cyclb_{m/2}),
\end{equation}
where $p_J$ and $q_{J}$ satisfy the conditions,
\begin{eqnarray}
&&
q_{\tilde{J}}-q_J\equiv p_J+p_{J+m/2},
\nn \\
&&
q_{J+m/2}\equiv q_J, \quad \mod n,
\label{pq cond tIIAB}
\end{eqnarray}
where $\tilde{J}$ is defined as, 
\begin{eqnarray}
&&\tilde{J}=\left\{
  \begin{array}{@{\,}ll}
    J^{\prime} & \mbox{($0\le J \le m/2-1$)} \\
    J^{\prime}+m/2 &\mbox{($m/2\le J \le m-1$)}
  \end{array}
\right. \nn \\
&&
J^{\prime} \equiv J+1 \quad  (\mbox{mod}\,\,\, m/2). \nn
\end{eqnarray}
\end{itemize}
The solution for the fermionic fields are obtained by $\bZ_2$ extension.
In the following, we we will solve the consistency conditions for $I_A$,
$II_A$ and $\widetilde{II}_A$.

\subsubsection{Fermionic partition function for $I_A$}

The periodicity and boundary condition for the $nm$ fermionic fields
which corresponds to the $I_A$-type bosonic solution is,
\begin{eqnarray}
h &=&
{\rm diag}(A_0\cycl_n,\cdots ,A_{m-1}\cycl_n),
\nn \\
g &=&
{\rm diag}(B_0\cycl_n^{p_0},\cdots ,B_{m-1}\cycl_n^{p_{m-1}})\cdot \cyclb_m,
\nn \\
f_1 &=&
{\rm diag}(F_0^{(1)}\cycl_n^{q_0}\inv_n,\cdots ,
F_{m-1}^{(1)}\cycl_n^{q_{m-1}}\inv_n),
\nn \\
f_2 &=&
{\rm diag}(F_0^{(2)}\cycl_n^{q_0+1}\inv_n,\cdots ,
F_{m-1}^{(2)}\cycl_n^{q_{m-1}+1}\inv_n),
\label{IA cond F}
\end{eqnarray}
where $A_{J}$, $B_{J}$ and $F_{J}^{(i)}$($i=1,2$) is the $\bZ_2$
twists as,
\begin{eqnarray}
A_{J} &=&
\diag\left(\be{\frac{\alpha_{0,J}}{2}},\cdots ,
\be{\frac{\alpha_{n-1,J}}{2}}\right),
\nn \\
B_{J} &=&
\diag\left(\be{\frac{\beta_{0,J}}{2}},\cdots ,
\be{\frac{\beta_{n-1,J}}{2}}\right),
\nn \\
F_{J}^{(i)} &=&
\diag\left(\be{\frac{\phi_{0,J}^{(i)}}{2}},\cdots
,\be{\frac{\phi_{n-1,J}^{(i)}}{2}}\right),
\nn \\
&&
J=0,\cdots ,m-1, \quad i=1,2.
\end{eqnarray}
$\alpha_{I,J}$, $\beta_{I,J}$ and $\phi_{I,J}^{(i)}$ take value
$0$ or $1$.
From the fermionic consistency conditions
(\ref{h cond A}), (\ref{hg cond A}) and the left-right symmetry
(\ref{LR symmetry F}),
$\alpha_{I,J}$, $\beta_{I,J}$ and $\phi_{I,J}^{(i)}$ are constrained as,
\begin{eqnarray}
&&
\alpha_{I,J} = \phi_{I,J}^{(1)} -\phi_{I+1,J}^{(2)},
\nn \\ &&
\beta_{I,J} + \alpha_{I+p_J,J+1} = \alpha_{I,J} + \beta_{I+1,J},
\nn
\\
&&
\phi_{I,J}^{(1)} + \phi_{n-1+q_J-I,J}^{(1)} = \epsilon_{I,J}, \nn\\
&&
\phi_{I,J}^{(2)} + \phi_{n+q_J-I,J}^{(2)} = \epsilon_{I,J},
\label{rel IA}
\end{eqnarray}
where these equations are defined mod $2$.
$\epsilon_{I,J}$ is the matrix element of $\cE$
in (\ref{LR symmetry F})
which is defined as,
\begin{eqnarray}
\cE  &=&
\diag\left(
E_0, \cdots ,E_{m-1}
\right) \nn \\
E_{J} &=&
\diag\left(\be{\frac{\epsilon_{0,J}}{2}}, \cdots
,\be{\frac{\epsilon_{n-1,J}}{2}}
\right),
\end{eqnarray}
where $\epsilon_{I,J}$'s take value 0 or 1.
The constraint for $p_J$ and $q_J$ can be obtained from (\ref{fg cond A}),
\begin{equation}
q_{J+1}-q_J\equiv 2p_J, \quad (\mbox{mod}\,\,\, n).
\end{equation}
If we sum up $J$, we obtain the following relation,
\begin{eqnarray}
2p &\equiv& 0, \quad (\mbox{mod}\,\,\, n), \\
p &\equiv& \sum_{J=0}^{m-1} p_J,
 \nn
\end{eqnarray}
where $p$ is also defined mod $n$.
From the condition (\ref{rel IA}), we obtain the following relation
\begin{eqnarray}
\alpha_{J} &=& \alpha ,\quad J=0,\cdots ,m-1, \\
\alpha_{J} &\equiv& \sum_{I=0}^{n-1}\alpha_{I,J}.
\nn
\end{eqnarray}

Under these constraints,
the action of $h$ and $g$ is diagonalized as torus case.
The mode expansion for the diagonalized field $\psi^{a,J}$
with respect to $h$ is,
\begin{eqnarray}
\psi^{a,J}_{L}(\sigma^0,\sigma^1) &=&
i^{-1/2}
\sum_{r\in \bZ}\psi^{a,J}_{r+\frac{a}{n}+\frac{\alpha_J}{2n}}
{\rm exp}\left[i(r+\frac{a}{n}+\frac{\alpha_J}{2n})w
\right], \nn \\
\psi^{a,J}_{R}(\sigma^0,\sigma^1) &=&
(f_2 \cdot \psi_{L})^{a,J}(\sigma^0,\sigma^1), \nn \\
&=&
\be{-\frac{a+\alpha/2}{n}(n+q_J)
+\frac{\phi_{0,J}^{(2)}}{2}
+\sum_{K=0}^{n+q_J-1}\frac{\alpha_{K,J}}{2}} \nn \\
&&
\times i^{-1/2}\sum_{r\in \bZ}
\bar{\psi}^{a,J}_{r-\frac{a}{n}-\frac{\alpha}{2n}}
{\rm exp}\left[i(r-\frac{a}{n}-\frac{\alpha}{2})w\right].
\label{mode IA}
\end{eqnarray}
The action of $g$ on the oscillators is evaluated as,
\begin{equation}
\left(g\cdot \psi_{r+\frac{a}{n}+\frac{\alpha}{2n}} \right)^{a,J}=
\be{\frac{\beta_{0,J}}{2}+\frac{(a+\alpha/2)p_J}{n}
-\sum_{K=0}^{p_J-1}\frac{\alpha_{K,J+1}}{2}}
\psi^{a,J+1}_{r+\frac{a}{n}+\frac{\alpha}{2n}}
\end{equation}

Using the oscillators and the eigenvalues of $g$, we can evaluate
the fermionic partition function for this condition.
The calculation is the same as that of the torus case
but the value of $p$ is restricted to $2p\equiv 0$ (mod $n$).
\begin{itemize}
\item Odd $n$ case:

For odd $n$ case, only $p=0$ is admitted.
The fermionic partition function for this case is,
\begin{eqnarray}
&& Z^{-}\left(\tau_{n,m,0}\right), \\
&&\tau_{n,m,0} = \frac{m}{n}\tau. \nn
\end{eqnarray}
This can be interpreted as the partition function for the large annulus.

\item Even $n$ case:

For even $n$ case, $p=0$ and $p=n/2$ are admitted.
The fermionic partition function for $p=0$ is same as above
and the fermionic partition function for $p=n/2$ 
becomes,
\begin{eqnarray}
Z^{-}\left(\tau_{n,m,0}+1/2
\right)
\end{eqnarray}
This can be interpreted as the partition function for the large M\"obius
strip.
\end{itemize}

Thus from $I_A$ type solution,
we obtain the annulus and M\"obius strip amplitude for the
long string.

\subsubsection{Fermionic partition function for $II_A$}

For even $m$, we have the solution which corresponds to
the bosonic $II_A$-type.
The solution has the same form as $I_A$ case for $h$ and $g$.
The solution for $f_1$ and $f_2$ is,
\begin{eqnarray}
f_1 &=&
{\rm diag}(F_{0}^{(1)}\cycl_n^{q_0}\inv_n,\cdots
,F_{m-1}^{(2)}\cycl_n^{q_{m-1}}\inv_n)
\cdot \cyclb_m^{m/2}, \nn \\
f_2 &=&
{\diag}(F_{0}^{(2)}\cycl_n^{q_0+1}\inv_n,\cdots
,F_{m-1}^{(2)}\cycl_n^{q_{m-1}}\inv_n)
\cdot \cyclb_m^{m/2}.
\end{eqnarray}

The constraints for $\alpha_{I,J}$, $\beta_{I,J}$ and $\phi_{I,J}^{(i)}$
which come from the fermionic consistency conditions
(\ref{h cond A}) and (\ref{hg cond A})
is same as that of $I_A$-type solution.
The constraints which come from the left-right symmetry
(\ref{LR symmetry F}) is
\begin{eqnarray}
\phi_{I,J}^{(1)}+\phi_{n-1+q_J-I,J+k}^{(1)} &=& \epsilon_{I,J}, \nn \\
\phi_{I,J}^{(2)}+\phi_{n+q_J-I,J+k}^{(2)} &=& \epsilon_{I,J},
\end{eqnarray}
where $k=m/2$.
The constrints for $p_J$ and $q_J$ is,
\begin{eqnarray}
q_{J+1}-q_J&\equiv& p_J+p_{J+m/2}, \quad (\mbox{mod}\,\,\, n), \nn
 \\
q_{J}&=&q_{J+k}.
\end{eqnarray}
Summing up $J$, we obtain the following relation,
\begin{equation}
\sum_{J=0}^{2k-1}p_J=0,
\end{equation}
where we denote $k\equiv m/2$.

Under these constraints, we will evaluate the eigenvalue for $h$ and $g$.
The mode expansion which satisfies this boundary condition can be
obtained as $I_A$ case,
\begin{eqnarray}
\psi^{a,J}_{L}(\sigma^0,\sigma^1) &=&
i^{-1/2}
\sum_{r\in \bZ}\psi^{a,J}_{r+\frac{a}{n}+\frac{\alpha_J}{2n}}
{\rm exp}\left[i(r+\frac{a}{n}+\frac{\alpha_J}{2n})w
\right],   \nn \\
\psi^{a,J}_{R}(\sigma^0,\sigma^1)
&=&
\be{-\frac{a+\alpha_J/2}{n}(n+q_J)
+\frac{\phi_{0,J}^{(2)}}{2}
+\sum_{K=0}^{n+q_J-1}\frac{\alpha_{K,J+k}}{2}} \nn \\
&&
\times i^{-1/2}\sum_{r\in \bZ}
\bar{\psi}^{a,J+k}_{r-\frac{a}{n}-\frac{\alpha_{J+k}}{2n}}
{\rm exp}\left[i(r-\frac{a}{n}-\frac{\alpha_{J+k}}{2})w\right],
\label{mode IIA}
\end{eqnarray}
where we used the relation 
$\alpha_J=\alpha$ for all $J=0, \cdots ,m-1$
as $I_A$ case.
The action of $g$ is the same as $I_A$ case.

Thus the partition function can be calculated as the torus with $p=0$
as,
\begin{equation}
Z^{-}(2\tau_{n,k,0}).
\end{equation}
This partition function can be interpreted as the Klein bottle partition
function for the long string.

\subsubsection{Fermionic partition function for $\widetilde{II}_A$}

For even $m$, we have another solution which corresponds to the bosonic
$\widetilde{II}_A$-type solution.
This solution has same $h$, $f_1$ and $f_2$ as the $II_A$-type
solution.
The solution for $g$ is
\begin{equation}
g =
\diag(B_{0}\cycl_n^{p_0}, \cdots ,B_{m-1}\cycl_n^{p_{m-1}})\cdot
\diag(\cyclb_{m/2},\cyclb_{m/2}).
\end{equation}
From the consistency condition, $p$'s and $q$'s should satisfy the
constraints as,
\begin{eqnarray}
q_{\tilde{J}}-q_{J} \equiv p_J+p_{J+m/2},\quad
q_{J+m/2} &=& q_{J},
\label{tIIA constraint}
\end{eqnarray}
where the $\tilde{J}$ is defined as,
\begin{eqnarray}
&&\tilde{J}=\left\{
  \begin{array}{@{\,}ll}
    J^{\prime} & \mbox{($0\le J \le m/2-1$)} \\
    J^{\prime}+m/2 &\mbox{($m/2\le J \le m-1$)}
  \end{array}
\right. \nn \\
&&
J^{\prime} \equiv J+1 \quad  (\mbox{mod}\,\,\, m/2). \nn
\end{eqnarray}
By the consistency condition (\ref{hg cond A}),
we have the constraint for $\alpha_{I,J}$
and $\beta_{I,J}$ as,
\begin{equation}
\alpha_{I,J}+\beta_{I+1,J}=\beta_{I,J}+\alpha_{I+p_J,\tilde{J}}.
\label{ab cond tIIA}
\end{equation}
The other constraints are same as $II_A$ case.

The mode expansions for the fermionic fields are same as $II_A$ case
(\ref{mode IIA}).
In $\widetilde{II}_A$ case, we should define
\begin{eqnarray}
\alpha_{L} &\equiv&  \alpha_{J}, \quad J=0,1,\cdots,k-1 ,\nn \\
\alpha_{R} &\equiv&  \alpha_{J}, \quad J=k,k+1,\cdots,m-1 ,
\end{eqnarray}
where we denote $k\equiv m/2$.
$\alpha_{L}$ and $\alpha_{R}$ do not depend on $J$ by the constraint
(\ref{ab cond tIIA}) and take independent value.

The eigenvalue of $g$
is evaluated independently
for the diagonalized field
$\Psi^{a,J}$ ($J=0, \cdots ,k-1$)  and $\Psi^{a,J}$ ($J=k, \cdots ,m-1$).

From the consistency condition (\ref{fg cond A}),
\begin{eqnarray}
&&q_{\tilde{J}}-q_J\equiv p_{J+k}+p_J, \quad
(\mbox{mod}\,\,\, n),
\nn \\
&&q_{J}=q_{J+k}.
\end{eqnarray}
Therefore $p_L\equiv \sum_{J=0}^{k-1}p_J$ and
$p_R\equiv \sum_{J=k}^{m-1}p_J$ are related as,
\begin{equation}
p_L=-p_R\equiv p  \quad (\mbox{mod}\,\,\, n).
\end{equation}

The oscillator part of
the partition function for $\Psi^{I,J}$ is evaluated as,
\begin{eqnarray}
&&
\prod_{r \in \bN-\alpha_L/2}\left(
1-\be{\frac{\beta_L^{\{p\}}-1}{2}}\be{\frac{k\tau-p}{n}r}
\right)
\prod_{s \in \bN-\alpha_R/2}
\left(
1-\be{\frac{\beta_R^{\{p\}}-1}{2}}\be{\frac{k\tau+p}{n}s}
\right),\nn \\
&&
\beta_{L}^{\{p\}}\equiv \sum_{J=0}^{k-1}(\beta_{0,J}
+\sum_{K=0}^{p_J-1}\alpha_{K,J+1}), \quad
\beta_{R}^{\{p\}}\equiv \sum_{J=k}^{m-1}(\beta_{0,J}
+\sum_{K=0}^{p_J-1}\alpha_{K,J+1}).
\end{eqnarray}

Gathering various factors, we obtain the partition function for
$\widetilde{II}_A$-type solutiuon,
\begin{equation}
Z^{-}(\tau_{n,k,p})Z^{-}(\bar{\tau}_{n,k,p})^{*},
\end{equation}
Thus we obtained the torus partition function for the long string.

\subsection{Partition function for M\"obius strip diagram}

The partition function for the unoriented open superstring is defined as
that of the M\"obius strip,
\begin{equation}
{\rm Tr}_{\cH_{f_1,f_2}}(\Omega g \be{\tau L_0^{{\rm open}}}),
\end{equation}
where $\Omega$ is the orientation flip for the open string.

From the path-integral  consistency condition for this partition function,
the invariance of the Hilbert space under the action of $\Omega g$
is necessary.
The action of $\Omega g$ on the left- and right-moving fermionic field
is,
\begin{eqnarray}
\Omega g\cdot\Psi_L(\sigma^1)&=&
g\cdot\Psi_R(\pi-\sigma^1), \nn \\
\Omega g\cdot\Psi_R(\sigma^1)&=&
\cC g\cdot\Psi_L(\pi-\sigma^1).
\end{eqnarray}
Since
these fermionic fields satisfy the open string boundary condition
(\ref{boundary cond F}),
the boundary condition for $\Omega g$ acted fermionic field is,
\begin{eqnarray}
\Omega g\cdot\Psi_R(0)&=&
\cC g\cdot\Psi_L(\pi), \nn \\
&=&\cC g f_2^{-1}\cdot\Psi_R(\pi), \nn \\
&=&\cC g f_2^{-1}g^{-1}\cdot
\left(\Omega g\cdot\Psi_R(0)\right),
\end{eqnarray}
where we denoted the boundary condition at $\sigma^1=0$,
the boundary condition at $\sigma^1=\pi$ can be evaluated as above.
As
the boundary condition must be invariant under the action of $\Omega g$
for the consistency of the Hilbert space, we need the constraint,
\begin{equation}
f_1=\cC g f_2^{-1}g^{-1}, \quad f_2=\cC g f_1^{-1}g^{-1}.
\end{equation}
The twist along the time direction can be obtained as,
\begin{equation}
h=f_2^{-1}f_1=gf_1g^{-1}\cC^{-1}f_1.
\label{h cond MS}
\end{equation}
If these $h$ and $g$ satisfy the consistency condition
for the unoriented world sheet,
\begin{equation}
ghg^{-1}=h^{-1},
\end{equation}
$\cC$, $f_1$, $f_2$, and $g$ should be constrained as,
\begin{equation}
gf_1^{2}g^{-1}=\cC f_1\cC f_1.
\label{cfg cond MS}
\end{equation}

In terms of the path integral on the M\"obius strip diagram,
the fermionic fields in the Hilbert space
$\cH_{f_1,f_2}$ must satisfy the following conditions at
$\sigma^1=0,\pi/2$,
\begin{eqnarray}
\Psi_{R}(\sigma^0,0)&=& f_1\cdot\Psi_L(\sigma^0,0), \nn \\
\Psi_{R}(\sigma^0+\pi,\pi/2)&=& \cC g\cdot\Psi_L(\sigma^0,\pi/2), \nn \\
\Psi_{L}(\sigma^0+\pi,\pi/2)&=&  g\cdot\Psi_R(\sigma^0,\pi/2).
\end{eqnarray}
The boundary condition at $\sigma^1=\pi/2$ is the cross-cap condition.
It is consistent if the following condition is satisfied,
\begin{eqnarray}
[g,\cC]=0.\label{cg cond MS}
\end{eqnarray}

In order to consider the mode expansion, we introduce the chiral field
$\Psi$ on the double cover of the M\"obius diagram (Figure 4).\cite{Pradisi-Sagnotti}

 \begin{figure}[ht]
  \centerline{\epsfxsize=5cm \epsfbox{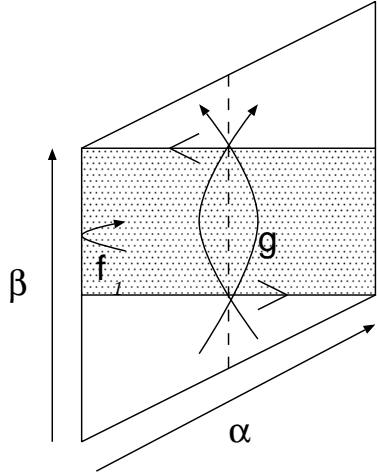}}
  \vskip 3mm
  \caption{Double cover for M\"obius  strip diagram}
 \end{figure}

The chiral field satisfies the following periodicity,
\begin{eqnarray}
\Psi(\sigma^0+\pi,\sigma^1+\pi)&=&\alpha\cdot\Psi(\sigma^0,\sigma^1), \nn \\
\Psi(\sigma^0+2\pi,\sigma^1)&=&\beta\cdot\Psi(\sigma^0,\sigma^1),
\quad \alpha,\beta \in \Gamma\times \bZ_2.
\end{eqnarray}
By the above boundary conditions,
the chiral field $\Psi$ and the fields in $\cH_{f_1,f_2}$ can be
identified as,
\begin{eqnarray}
\Psi_L(\sigma^0,\sigma^1) &=& \Psi(\sigma^0,\sigma^1), \nn \\
\Psi_R(\sigma^0,\sigma^1) &=& g^{-1}\cdot \Psi(\sigma^0+\pi,\pi-\sigma^1).
\end{eqnarray}
The periodicities can be written by the boundary twists as,
\begin{equation}
\alpha=gf_1,\quad \beta=\cC g^2.
\end{equation}
The periodicity and the boundary condition along $\sigma^1=0$
implies,
\begin{equation}
[f_1,\beta]=[f_1,\cC g^2]=0.
\label{inv cond MS}
\end{equation}
Thus we specfied the unoriented open string Hilbert space from the
boundary conditions and obtained the consistent boundary conditions.

In the bosonic case\cite{Matsuo-Fuji},
the solution of the periodicity and the boundary
conditions for the irreducible set of the
permutation group are calssified as $I_M$, $II_M$ and $\widetilde{II}_M$.
\begin{itemize}
\item $I_M$ type solution:
\begin{eqnarray}
h&=&\diag(\overbrace{\cycl_n,\cdots,\cycl_n}^{m}),
\nn \\
g&=&\diag(\cycl_n^{p_0}\inv_n,\cdots,\cycl_n^{p_{m-1}}\inv_n)\cdot\cyclb_m,
\nn \\
f_1&=&\diag(\cycl_n^{q_0}\inv_n,\cdots,\cycl_n^{q_{m-1}}\inv_n),
\end{eqnarray}
where $p_J,q_J$ take their values in $0,1,\cdots,n-1$ (mod $n$) and 
$p_J$ and $q_J$ satisfy the condition, 
\begin{equation}
q_{J+1}+q_J\equiv 2p_J-1 \mod n.
\label{pq cond IMB}
\end{equation}

\item $II_M$ type solution:

For even $m$, there is a solution which has the same form of $h$ and $g$ as
$I_M$ case and $f_1$ takes the form as,
\begin{equation}
f_1=\diag(\cycl_n^{q_0}\inv_n,\cdots,\cycl_n^{q_{m-1}}\inv_n)
\cdot(\cyclb_{m/2})^{m/2}
\end{equation}
where $p_J$ and $q_J$ satisfy the conditions,
\begin{eqnarray}
&&
q_{J+1}+q_J\equiv p_J+p_{J+m/2}-1, \nn \\
&&
q_{J+m/2}\equiv q_{J}, \quad \mod n.
\label{pq cond IIMB}
\end{eqnarray}

\item $\widetilde{II}_M$ type solution:

For even $m$, there is another solution which has the same form of 
$h$ and $f_1$ as $II_M$ case and $g$ takes the form as,
\begin{equation}
g=\diag(\cycl_n^{p_0}\inv_n,\cdots,\cycl_n^{p_{m-1}}\inv_n)\cdot
\diag(\cyclb_{m/2},\cyclb_{m/2}),
\end{equation}
where $p_J$ and $q_{J}$ satisfy the conditions,
\begin{eqnarray}
&&
q_{\tilde{J}}+q_J\equiv p_J+p_{J+m/2}-1,
\nn \\
&&
q_{J+m/2}\equiv q_J, \quad \mod n,
\label{pq cond tIIMB}
\end{eqnarray}
where $\tilde{J}$ is defined as, 
\begin{eqnarray}
&&\tilde{J}=\left\{
  \begin{array}{@{\,}ll}
    J^{\prime} & \mbox{($0\le J \le m/2-1$)} \\
    J^{\prime}+m/2 &\mbox{($m/2\le J \le m-1$)}
  \end{array}
\right. \nn \\
&&
J^{\prime} \equiv J+1 \quad  (\mbox{mod}\,\,\, m/2). \nn
\end{eqnarray}
\end{itemize}
The solution for the fermionic boundary conditions can be
obtained by $\bZ_2$-extending the bosonic solution and
then solving the consistency condition for the $\bZ_2$-twists.

\subsubsection{Fermionic partition function for $I_M$}

The periodicity and the boundary condition for $nm$ fermionic fields
which corresponds to the $I_M$-type bosonic solution is
\begin{eqnarray}
h&=&\diag(A_0\cycl_n,\cdots ,A_{m-1}\cycl_n), \nn \\
g&=&\diag(B_{0}\cycl_n^{p_{0}}\inv_n,\cdots ,
B_{m-1}\cycl_n^{p_{m-1}}\inv_n)\cdot \cyclb_m, \nn  \\
f_1&=&\diag(F_0\cycl_n^{q_0}\inv_n,\cdots ,F_{m-1}\cycl_n^{q_{m-1}}\inv_n),
\nn \\
\cC&=&\diag(C_0, \cdots ,C_{m-1}), \nn \\
C_J&\equiv&\diag\left(\be{\frac{c_{0,J}}{2}}, \cdots
,\be{\frac{c_{n-1,J}}{2}}
\right), \quad J=0,\cdots ,m-1.
\end{eqnarray}
We will find the consistency condition for $\alpha$'s, $\beta$'s,
$\phi$'s and $c$'s.
From the consistency condition for the M\"obius strip diagram
(\ref{cg cond MS}),(\ref{h cond MS}),(\ref{inv cond MS}),(\ref{cfg cond
MS}),
we obtain the following relations,
\begin{eqnarray}
&&
c_{I,J}=c_{n-1-I+p_J,J+1}, \nn \\
&&
\alpha_{I,J}=\phi_{I,J}+\phi_{I+p_{J-1}-q_J,J-1}
-\beta_{I+p_{J-1}-q_J,J-1} \nn \\
&& \quad \quad\quad
+\beta_{n-1+q_{J-1}+q_J-p_{J-1}-I,J-1}+c_{n-1-I+q_J,J}, \nn \\
&&
2p_{J-1}-q_{J-1}-q_{J}\equiv 1 \quad \mod{n}, \nn \\
&&
\phi_{I,J}+\phi_{I+P_{J+1}-p_{J},J+2}-(c_{I,J}-c_{n-1+q_J-I,J})\nn \\
&&\quad
-(\beta_{I,J}-\beta_{n-1+q_J-I,J})+(\beta_{I+p_J-q_J,J+1}-\beta_{n-1+p_J-I,J
+1})
=0,\nn \\
&&
\phi_{n-1+p_{J-1}-I,J-1}+\phi_{I+q_{J-1}-p_{J-1},J-1}, \nn \\
&& \quad
=\phi_{I,J}+c_{I,J}+\phi_{n-1+q_J-I,J}+c_{n-1+q_J-I,J}.\label{cond IM}
\end{eqnarray}
If we sum up $I$ in the above constraints,
we obtain the following relations,
\begin{eqnarray}
&&
\alpha_{J}=\phi_J-\phi_{J-1}+c_J, \quad
\phi_{J}=\phi_{J+2}, \quad
c_{J}=c_{J+1}, \\
&&
\alpha_{J}\equiv\sum_{I=0}^{n-1}\alpha_{I,J}, \quad
c_J\equiv \sum_{I=0}^{n-1}c_{I,J}, \quad
\phi_{J}\equiv\sum_{I=0}^{n-1}\phi_{I,J}.
\nn
\end{eqnarray}
Using these relations, we get
\begin{equation}
\alpha_{J}\equiv \alpha, \quad J=0,\cdots ,m-1.
\end{equation}

Under these constraints, the action of $\Omega g$ on the oscillator 
$\psi_{r+\frac{a}{n}+\frac{\alpha}{2n}}^{a,J}$
in the mode expansion (\ref{mode IA}) is evaluated as,
\begin{eqnarray}
&&
(\Omega g\cdot \psi)_{r+\frac{a}{n}+\frac{\alpha}{2n}}^{a,J} =
\nn \\
&&
\be{\frac{a+\alpha/2}{n}(p_J-q_J)+\frac{1}{2}\left(
\phi_{0,J}+\beta_{n-1+q_J,J}-\sum_{K=0}^{p_J-q_J-1}\alpha_{K,J+1}
\right)-\frac{a+\alpha/2}{2n}}
\nn \\
&&
\times
(-1)^{r}\psi_{r+\frac{a}{n}+\frac{\alpha}{2n}}^{a,J+1}.
\end{eqnarray}
The eigenvalue for $\Omega g$ is 
evaluated as torus case by diagonalizing $J$.

Thus the oscillator part of the partition function for $I_M$ case
is calculated as,
\begin{eqnarray}
&&\prod_{a=0}^{n-1}\prod_{b=0}^{m-1}Z_{a,b}
=
\prod_{r=1}^{\infty}\prod_{a=0}^{n-1}
\left(
1-(-1)^{mr}\be{\frac{\beta^{\{p,q\}}}{2}}
\be{\frac{a+\alpha/2}{n}\cdot\frac{p}{2}}
\be{m\tau(r-\frac{a}{n}-\frac{\alpha}{2n})}
\right), \nn \\
&& p\equiv \sum_{J=0}^{m-1}(2p_J-2q_J-1), \quad
\beta^{\{p,q\}}\equiv\sum_{J=0}^{m-1}\left(\phi_{0,J}+\beta_{n-1+q_J,J}
-\sum_{K=0}^{p_J-q_J-1}\alpha_{K,J+1}\right).
\end{eqnarray}

The constraint (\ref{cond IM}) implies $p \equiv 0$ (mod $n$).
However we need to evaluate it in mod $2n$ to get the accurate phase factor.
So the partition function can be classified as follows.
\begin{itemize}
\item $n$:odd and $m$:odd

In this case, only $p\equiv n$ (mod $2n$) can be admitted.
Using the relation  $(-1)^{ms}=(-1)^s=(-1)^{ns}$ and
gathering all factors, we obtain the partition function,
\begin{equation}
Z(n,m|\tau)
=Z^{-}(\tau_{n,m,0}+\frac{1}{2}).
\end{equation}
In this case, we can interpret this result as the partition function
for the large M\"obius strip.

\item $n$:odd and $m$:even

In this case, only $p\equiv 0$ (mod $2n$) can be admitted.
Using $(-1)^{ms}=1$ and
gathering all factors, we obtain the partition function,
\begin{equation}
Z(n,m|\tau)
=Z^{-}(\tau_{n,m,0}).
\end{equation}
In this case, we can interpret this result as the partition function
for the large annulus.

\item $n$:even and $m$:even

In this case, $p\equiv 0,n$ (mod $2n$) can be admitted.
The partition function can be calculated as above.
\begin{enumerate}
\item $p\equiv 0$ case

If we use $(-1)^{ms}=1$, the partition function becomes,
\begin{equation}
Z(n,m|\tau)
=Z^{-}(\tau_{n,m,0}).
\end{equation}
Thus we obtained the partition function for the large
annulus.

\item $p\equiv n$ case

If we use $(-1)^{ms}=1=(-1)^{ns}$, the partition function becomes,
\begin{equation}
Z(n,m|\tau)
=Z^{-}(\tau_{n,m,0}+1/2).
\end{equation}
Thus we obtained the partition function for
the large M\"obius strip.

\end{enumerate}
\end{itemize}

\subsubsection{Fermionic partition function for $II_M$}

For even $m$, we have the solution which corresponds to
the bosonic $II_M$-type solution.
The solution has the same form as $I_M$ case for $h$ and $g$.
The solution for $f_1$ is
\begin{equation}
f_1 =
\diag(F_0\cycl_n^{q_0}\inv_n, \cdots ,F_{m-1}\cycl_n^{q_{m-1}}\inv_n)
\cdot \cyclb_m^{m/2}.
\end{equation}
In this case, the consistency condition becomes,
\begin{eqnarray}
&&
\alpha_{I,J}=
\phi_{I,J}+\phi_{I+p_{J+k-1}-q_J,J+k-1}-\beta_{I+p_{J+k-1}-q_J,J+k-1}\nn \\
&& \quad\quad\quad
+\beta_{n-1+q_{J+k-1}+q_J-p_{J+k-1}-I,J-1}
+c_{n-1-I+q_{J},J+k}, \nn \\
&&
p_{J-1}+p_{J+k-1}-q_{J+k-1}-q_{J} \equiv 1, \quad \mod{n},
\nn \\
&&
q_{J}=q_{J+k},
\nn \\
&&
\phi_{I,J}+\phi_{I+P_{J+1}-p_{J},J+2}-(c_{I,J}-c_{n-1+q_J-I,J+k})\nn \\
&&
-(\beta_{I,J}-\beta_{n-1+q_J-I,J+k})
+(\beta_{I+p_{J+k}-q_J,J+k+1}-\beta_{n-1+p_J-I,J+1})
=0,\nn \\
&&
\phi_{n-1+p_{J-1}-I,J-1}+\phi_{I+q_{J-1}-p_{J-1},J+k-1}, \nn \\
&& \quad
=\phi_{I,J}+c_{I,J}+\phi_{n-1+q_J-I,J+k}+c_{n-1+q_J-I,J+k}.\label{cond IIM}
\end{eqnarray}
where we denote $k=m/2$.
The condition $c_J$ is same as $I_M$ case.
If we sum up $I$ in the above constraints, we obtain the following
relations,
\begin{eqnarray}
&&
\alpha_J\equiv \phi_J + \phi_{J+k-1} +\beta_{J-1} -\beta_{J+k-1} +c_{J+k},
\nn \\
&&
\phi_J-\phi_{J+2}
\equiv (\beta_{J}-\beta_{J+k})-(\beta_{J+k+1}-\beta_{J+1})
+(c_J-c_{J+k}),\nn \\
&&
\beta_{J}\equiv \sum_{I=0}^{n-1}\beta_{I,J}.
\end{eqnarray}
Using these relations, we get the recursion relation,
\begin{equation}
\alpha_{J}\equiv \alpha_{J+k+1}.
\end{equation}
For odd $k$ case, this recursion relation is solved as,
\begin{equation}
\alpha_{R}\equiv \alpha_{0}= \alpha_{2} \cdots
\alpha_{2k-2}, \quad
\alpha_{L}\equiv \alpha_{1}= \alpha_{3} \cdots
\alpha_{2k-1}.
\end{equation}
For even $k$ case, this recursion relation is solved as,
\begin{equation}
\alpha\equiv \alpha_{J}, \quad J=0,\cdots , m-1.
\end{equation}

Under these constraints,
the action of $\Omega g$ on the oscillator in (\ref{mode IIA}) is
evaluated as,
\begin{eqnarray}
&&
(\Omega g\cdot \psi)_{r+\frac{a}{n}+\frac{\alpha_J}{2n}}^{a,J} =
\nn \\
&&
\be{\frac{a+\alpha_J/2}{n}(p_{J+k}-q_J)+\frac{1}{2}\left(
\phi_{0,J}+\beta_{n-1+q_J,J+k}-\sum_{K=0}^{p_{J+k}-q_J-1}\alpha_{K,J+k+1}
\right)-\frac{a+\alpha_J/2}{2n}}
\nn \\
&&
\times
(-1)^{r}\psi_{r+\frac{a}{n}+\frac{\alpha_J}{2n}}^{a,J+k+1}.
\end{eqnarray}
Since the value of $\alpha_J$ depends on $k$,
we classify the partition function as follows.

\begin{itemize}

\item $k$:even case

In this case, $\alpha_J=\alpha$.
By the constraint (\ref{cond IIM}), we have only
$p\equiv\sum_{J=0}^{m-1}(2p_{J+k}-2q_J-1)=0$.
Thus the oscillator part of the partition function is evaluated as,
\begin{eqnarray}
&&
\prod_{a=0}^{n-1}\prod_{b=0}^{m-1}Z_{a,b}=
\prod_{r \in\bN-\alpha/2}\left(
1-\be{\frac{\beta^{\{p,q\}}}{2}}\be{\frac{2k\tau}{n}r}
\right), \nn \\
&&
\beta^{\{p,q\}}\equiv \sum_{J=0}^{m-1}
\left(\phi_{0,J}+\beta_{n-1+q_J,J+k}
+\sum_{K=0}^{p_{J+k}-q_J-1}\alpha_{K,J+k+1}
\right).
\end{eqnarray}
Gathering all factors, we obtain the partition function for $II_M$ case,
\begin{equation}
Z(n,m|\tau)=Z^{-}(2\tau_{n,k,0}).
\end{equation}
Identifying $\psi^{a,J}$ as the left-mover for  $J$:odd and
as the right-mover for $J$:even, 
we can interpret this partition function as that of large Klein bottle.

\item $k$:odd case

In this case, we can take independent value for $\alpha_L$ and $\alpha_R$.
So the partition functions are evaluated for
odd and even $J$ independently.
Thus the oscillator part of the partition function is
evaluated as,
\begin{eqnarray}
&&
\prod_{a=0}^{n-1}\prod_{b=0}^{k-1}
Z_{a,b}^{J:{\rm odd}}Z_{a,b}^{J:{\rm even}}, \nn \\
&=&\prod_{r\in \bN -\alpha_L/2}\left(
1-\be{\frac{\beta^{\{p,q\}}_L}{2}}\be{\frac{k\tau-p}{n}r}
\right)
\prod_{r\in \bN -\alpha_R/2}\left(
1-\be{\frac{\beta^{\{p,q\}}_R}{2}}\be{\frac{k\tau+p}{n}r}
\right),\nn \\
&&
p\equiv\sum_{M=0}^{k-1}p_{2M+1+k}-\sum_{M=0}^{k-1}q_{2M+1}-k/2,
\nn \\
&&
\beta^{\{p,q\}}_L\equiv
\sum_{J:{\rm odd}}\left(\phi_{0,J}+\beta_{n-1+q_J,J+k}
-\sum_{K=0}^{p_{J+k}-q_J-1}\alpha_{K,J+k+1}
\right), \nn \\
&&
\beta^{\{p,q\}}_R\equiv \sum_{J:{\rm even}}\left(
\phi_{0,J}
+\beta_{n-1+q_J,J+k}-\sum_{K=0}^{p_{J+k}-q_J-1}\alpha_{K,J+k+1}
\right).
\end{eqnarray}
Gathering all factors, we obtain the partition
function,
\begin{equation}
Z(n,m,p|\tau)=Z^{-}(\tau_{n,k,p})Z^{-}(\bar{\tau}_{n,k,p})^{*}.
\end{equation}
This can be interpreted as the partition function for the large torus.
However, $p$ takes half-odd value.
So we call this diagram ``twisted'' torus.

\end{itemize}
\subsubsection{Fermionic partition function for $\widetilde{II}_M$}

For even $m$, we have another solution which corresponds to
the bosonic $\widetilde{II_M}$-type solution.
The solution has the same form as $II_M$ case for $h$ and $f_1$.
The solution for $g$ is
\begin{equation}
g=
\diag(B_0\cycl_n^{p_0}\inv_n,\cdots ,B_{m-1}\cycl_n^{p_{m-1}}\inv_n)
\cdot \diag(\cycl_{m/2},\cycl_m/2).
\end{equation}
The consistency condition becomes as,
\begin{eqnarray}
&&
c_{I,J}=c_{n-1+p_J-I,\tilde{J}},  \nn \\
&&
\alpha_{I,J}=\phi_{I,J}+\phi_{I+p_{\hat{J}+k}-q_J,\hat{J}+k}
-\beta_{I+p_{\hat{J}+k}-q_J,\hat{J}+k} \nn \\
&& \quad\quad\quad
+\beta_{n-1+q_{\hat{J}+k}+q_J-p_{\hat{J}+k}-I,\hat{J}}
+c_{n-1-I+q_{J},J+k}, \label{a cond tIIM} \nn \\
&&
p_{J}+p_{J+k}-q_{J+k}-q_{\tilde{J}} \equiv 1, \quad \mod{n},
\label{pq cond tIIM} \nn \\
&&
q_{J}=q_{J+k},
\label{q cond tIIM} \nn \\
&&
\phi_{I,J}+\phi_{I+P_{\tilde{J}}-p_{J},\tilde{\tilde{J}}}-(c_{I,J}-c_{n-1+q_
J-I,J+k})\nn \\
&&
-(\beta_{I,J}-\beta_{n-1+q_J-I,J+k})
+(\beta_{I+p_{J+k}-q_J,\tilde{J}+k}-\beta_{n-1+p_J-I,\tilde{J}})
=0,\label{b cond tIIM} \nn \\
&&
\phi_{n-1+p_{\hat{J}}-I,\hat{J}}+\phi_{I+q_{\hat{J}}-p_{\hat{J}},\hat{J}+k},
\nn \\
&& \quad
=\phi_{I,J}+c_{I,J}+\phi_{n-1+q_J-I,J+k}+c_{n-1+q_J-I,J+k},\label{cond tIIM}
\end{eqnarray}
where  we denoted as $k=m/2$ and defined $\hat{J}$ as,
\begin{eqnarray}
&&\hat{J}=\left\{
  \begin{array}{@{\,}ll}
    J^{\prime\prime} & \mbox{($0\le J \le m/2-1$)} \\
    J^{\prime\prime}+m/2 &\mbox{($m/2\le J \le m-1$)}
  \end{array}
\right. \nn \\
&&
J^{\prime\prime} \equiv J-1 \quad  (\mbox{mod}\,\,\, m/2). \nn
\end{eqnarray}
As $II_M$ case, we can find the following relations.
For even $k$,
\begin{eqnarray}
\alpha_{R}&\equiv&
\alpha_{0}=\alpha_{k+1}=\alpha_{2}=\cdots =\alpha_k, \nn \\
\alpha_{L}&\equiv&
\alpha_{1}=\alpha_{k+2}=\alpha_{3}=\cdots =\alpha_{k+1}.
\end{eqnarray}
For odd $k$,
\begin{equation}
\alpha\equiv\alpha_{J}, \quad J=0,\cdots ,m-1.
\end{equation}

The action of $\Omega g$ on the oscillator
$\psi^{a,J}_{r+a/n+\alpha/2n}$ is written as,
\begin{eqnarray}
&&
(\Omega g\cdot \psi)_{r+\frac{a}{n}+\frac{\alpha_J}{2n}}^{a,J} =
\nn \\
&&
\be{\frac{a+\alpha_J/2}{n}(p_{J+k}-q_J)+\frac{1}{2}\left(
\phi_{0,J}+\beta_{n-1+q_J,J+k}-\sum_{K=0}^{p_{J+k}-q_J-1}\alpha_{K,\tilde{J}
+k}
\right)-\frac{a+\alpha_J/2}{2n}}
\nn \\
&&
\times
(-1)^{r}\psi_{r+\frac{a}{n}+\frac{\alpha_J}{2n}}^{a,\tilde{J}+k}.
\end{eqnarray}
Since the value of $\alpha_J$ depends on $k$, we classify the partition
function as follows.

\begin{itemize}
\item $k$:odd case

In this case, $\alpha\equiv \alpha_J$.
From the constraint (\ref{pq cond tIIM}), only $p=0$ can be admitted.
The partition function can be calculated as in even $k$ case of $II_M$.
The result is,
\begin{equation}
Z(n,m|\tau)=Z^{-}(2\tau_{n,k,0}).
\end{equation}
This can be interpreted as the partition function on the large Klein
bottle.

\item $k$:even case

In this case, $\alpha_{L}$ and $\alpha_{R}$ can take independent values.
The partition function can be calculated as in odd $k$ case of $II_M$.
The oscillator part of the partition function can be written as,
\begin{eqnarray}
\prod_{a=0}^{n-1}\prod_{b=0}^{k-1}Z_{a,b}^{L}Z_{a,b}^{R}
&=&\prod_{r \in \bN -\alpha_{L}}\left(
1-\be{\frac{\beta^{\{p,q\}}_L}{2}}\be{\frac{k\tau-p}{n}r}
\right)\nn \\
&&\times
\prod_{r \in \bN -\alpha_{R}}\left(
1-\be{\frac{\beta^{\{p,q\}}_R}{2}}\be{\frac{k\tau+p}{n}r}
\right),\nn \\
p&\equiv&\sum_{M=0}^{k/2-1}(p_{2M}+p_{2M+k+1})-\sum_{M=0}^{k-1}q_M-k/2,
\end{eqnarray}
where $\beta^{\{p,q\}}_L$ and $\beta^{\{p,q\}}_R$ can be defined as
$II_M$ case.
Gathering all factors we obtain the partition function for
$\widetilde{II}_M$ as,
\begin{eqnarray}
Z(n,m,p|\tau)=Z^{-}(\tau_{n,k,p})Z^{-}(\bar{\tau}_{n,k,p})^{*}.
\end{eqnarray}
This can be interpreted as the partition function for the large torus.

\end{itemize}

\subsection{Partition function for Klein bottle diagram}

The partition function for the unoriented closed superstring is defined as,
\begin{equation}
{\rm Tr}_{\cH_{h}}
\left(\Omega g \be{\tau(L_0+\tilde{L}_0)}\right),
\label{trace MS}
\end{equation}
where $\Omega$ is the orientation flip operator for the closed string.
We will consider the consistent boundary condition for fermions.

From the path-integral  consistency condition for this partition function,
the invariance of the Hilbert space under the action of $\Omega g$
is necessary.
The actions of $\Omega g$ on the fermionic fields in the Hilbert space
$\cH_h$ are written as,
\begin{eqnarray}
\Omega g\cdot \Psi(\sigma^1)&=&g\cdot\tilde{\Psi}(2\pi-\sigma^1), \nn \\
\Omega g\cdot \tilde{\Psi}(\sigma^1)&=&g\cdot\Psi(2\pi-\sigma^1).
\end{eqnarray}
The periodicity of the $\Omega g$ acted fermionic field is evaluated as,
\begin{eqnarray}
\Omega g\cdot\Psi(\sigma^1+2\pi)&=&
g\cdot \tilde{\Psi}(-\sigma^1),\nn \\
&=&
gh^{-1}\cdot \tilde{\Psi}(2\pi-\sigma^1), \nn \\
&=&
gh^{-1}g^{-1}\cdot\left(\Omega g\cdot\Psi(\sigma^1+2\pi)\right).
\end{eqnarray}
As the Hilbert space $\cH_h$ must be invariant under the action of
$\Omega g$, we obtain the following constraint,
\begin{equation}
h^{-1}=ghg^{-1}. \label{hg cond KB}
\end{equation}

In terms of the path integral on the Klein bottle diagram,
the fermionic fields in the Hilbert space
$\cH_h$ must satisfy the following conditions at
$\sigma^1=0,\pi$,
\begin{eqnarray}
&&\Psi_{R}(\sigma^0+\pi,0)=\cC f_1\Psi_L(\sigma^0,0), \quad
\Psi_{L}(\sigma^0+\pi,0)= f_1\Psi_R(\sigma^0,0),
\label{LR1 cond KB}\\
&&\Psi_{R}(\sigma^0+\pi,\pi)=\cC f_2\Psi_L(\sigma^0,\pi), \quad
\Psi_{L}(\sigma^0+\pi,\pi)= f_2\Psi_R(\sigma^0,\pi).
\label{LR2 cond KB}
\end{eqnarray}
Both of these conditions are cross-cap conditions.
These conditions are consistent if the boundary twists satisty the
following conditions,
\begin{eqnarray}
&&
h=f_2^{-1}f_1, \nn \\
&&
[f_1,\cC]=[f_2,\cC]=0.
\end{eqnarray}

In order to consider the mode expansion,
we introduce the chiral field
$\Psi$ on the double cover of the 
Klein bottle diagram (Figure 5).\cite{Pradisi-Sagnotti}

 \begin{figure}[ht]
  \centerline{\epsfxsize=5cm \epsfbox{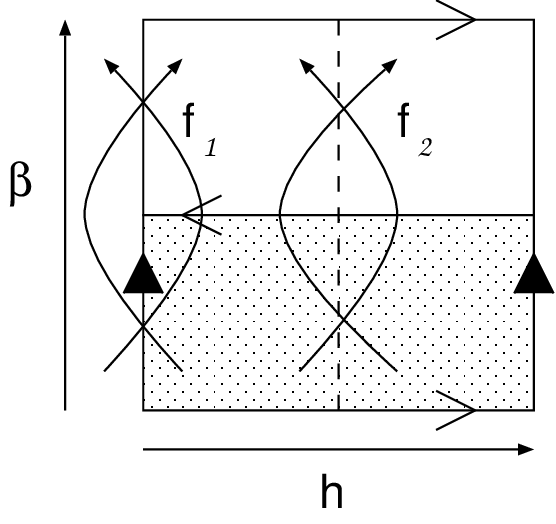}}
  \vskip 3mm
  \caption{Double cover for Klein bottle diagram}
 \end{figure}

This field has the following periodicities,
\begin{equation}
\Psi(\sigma^0,\sigma^1+2\pi) = h\cdot \Psi(\sigma^0,\sigma^1),
\quad
\Psi(\sigma^0+2\pi,\sigma^1) = \beta\cdot \Psi(\sigma^0,\sigma^1).
\end{equation}
We identify these fields as,
\begin{equation}
\Psi_L(\sigma^0,\sigma^1)=\Psi(\sigma^0,\sigma^1), \quad
\Psi_R(\sigma^0,\sigma^1)=\cC f_2\Psi(\sigma^0-\pi,2\pi-\sigma^1).
\end{equation}
The cross-cap conditions (\ref{LR1 cond KB}),(\ref{LR2 cond KB}),
the twist $\beta$ must be written in terms of the boundary twists as,
\begin{equation}
\beta=\cC f_1^2=\cC f_2^2. \label{beta cond KB}
\end{equation}
If we set $g=f_1$, two conditions (\ref{hg cond KB}) and
(\ref{beta cond KB}) are identical.

In the bosonic case\cite{Matsuo-Fuji}, the solution for $h$ and $g$ is
obtained for the irreducible set of the permutation group,
\begin{eqnarray}
h&=&\diag(\overbrace{\cycl_n,\cdots,\cycl_n}^m), \nn \\
g&=&\diag(\cycl_n^{p_0}\inv_n,\cdots,\cycl_n^{p_{m-1}}\inv_n)\cdot\cyclb_m.
\end{eqnarray}
The solution for the fermions can be obtained by $\bZ_2$ extention.
We will find the consistency condition for $\bZ_2$ twists and
evaluate the partition function in the following.

\subsubsection{Partition function for Klein bottle}

The solution for $nm$ left-moving fermionic fields $\Psi^{I,J}$
on the Klein bottle diagram is
\begin{eqnarray}
h&=&\diag(A_0\cycl_n,\cdots ,A_{m-1}\cycl_{n}), \nn \\
g&=&\diag(B_0\cycl_n^{p_0}\inv_n,\cdots ,B_{m-1}\cycl_n^{p_{m-1}}\inv_n)
\cdot\cyclb_m.
\end{eqnarray}
The solution for $nm$ right-moving fermionic fields $\tilde{\Psi}^{I,J}$ is
\begin{eqnarray}
h&=&\diag(\tilde{A}_0\cycl_n,\cdots ,\tilde{A}_{m-1}\cycl_{n}), \nn \\
g&=&\diag(\tilde{B}_0\cycl_n^{p_0}\inv_n,\cdots
,\tilde{B}_{m-1}\cycl_n^{p_{m-1}}\inv_n)
\cdot\cyclb_m,
\end{eqnarray}
where $\tilde{A}_J$ and $\tilde{B}_J$ are defined as,
\begin{eqnarray}
\tilde{A}_J&=&\diag\left(
\be{\frac{\tilde{\alpha}_{0,J}}{2}}, \cdots
,\be{\frac{\tilde{\alpha}_{m-1,J}}{2}}
\right), \nn \\
\tilde{B}_J&=&\diag\left(
\be{\frac{\tilde{\beta}_{0,J}}{2}}, \cdots
,\be{\frac{\tilde{\beta}_{m-1,J}}{2}}
\right). \nn
\end{eqnarray}
By the consistency condition (\ref{hg cond KB}),
the consistency condition for $\alpha$'s, $\beta$'s, $\tilde{\alpha}$'s
and $\tilde{\beta}$'s become,
\begin{eqnarray}
\alpha_{I,J}+\tilde{\alpha}_{n-2+p_J-I,J+1}&=&
\beta_{I,J}-\beta_{I+1,J}, \nn\\
\tilde{\alpha}_{I,J}+\alpha_{n-2+p_J-I,J+1}&=&
\tilde{\beta}_{I,J}-\tilde{\beta}_{I+1,J}.\label{ab cond KB}
\end{eqnarray}
Summing up $I$ in above relation, we obtain,
\begin{eqnarray}
&&\alpha_J=-\tilde{\alpha}_{J+1}, \quad \tilde{\alpha}_{J}=-\alpha_{J+1},
\nn\\
&&\tilde{\alpha}_J\equiv \sum_{I=0}^{n-1}\tilde{\alpha}_{I,J}.
\end{eqnarray}
For odd $m$ case, these relation can be solved as,
\begin{equation}
\alpha\equiv\alpha_{J}=-\tilde{\alpha}_{J}, \quad J=0, \cdots ,m-1.
\end{equation}
For even $m$ case,
\begin{equation}
\alpha_{L}\equiv \alpha_{2J} = -\tilde{\alpha}_{2J+1}, \quad
\alpha_{R}\equiv \alpha_{2J+1} = -\tilde{\alpha}_{2J}, \quad
J=0, \cdots ,m/2-1.
\end{equation}

Under these constraints, we will evaluate the eigenvalue for $\Omega g$.
$\Omega g$ acts on the fermionic field $\Psi,\tilde{\Psi}\in\cH_{h}$
as,
\begin{eqnarray}
(\Omega g\cdot
\Psi)^{I,J}=\be{\frac{\beta_{I,J}}{2}}\tilde{\Psi}^{n-1+p_J-I,J+1},\quad
(\Omega g\cdot
\tilde{\Psi})^{I,J}=\be{\frac{\tilde{\beta}_{I,J}}{2}}\Psi^{n-1+p_J-I,J+1}.
\end{eqnarray}
Therefore the action of $\Omega g$ on the oscillator in the mode expansion
(\ref{mode T2})
can be written as the M\"obius strip case,
\begin{eqnarray}
(\Omega g\cdot \psi)^{a,J}_{r+a/n+\alpha_J/2n}
&=&
\be{-\frac{a+\alpha_J/2}{n}(n-1+p_J)+\frac{\beta_{0,J}}{2}
-\sum_{K=0}^{n-2+p_J}\frac{\tilde{\alpha}_{K,J+1}}{2}}
\nn \\&&
\times
\tilde{\psi}^{-a,J+1}_{r+a/n-\tilde{\alpha}_{J+1}/2n},
\nn \\
(\Omega g\cdot \tilde{\psi})^{-a,J}_{r+a/n-\tilde{\alpha}_J/2n}
&=&
\be{-\frac{a-\tilde{\alpha}_J/2}{n}(n-1+p_J)+\frac{\tilde{\beta}_{0,J}}{2}
-\sum_{K=0}^{n-2+p_J}\frac{\alpha_{K,J+1}}{2}}
\nn \\&&
\times
\psi^{a,J+1}_{r+a/n+\alpha_{J+1}/2n}.
\end{eqnarray}

In order to evaluate the eigenvalue of $\Omega g$ on $\psi\tilde{\psi}$,
we need to find a combination of left- and right-movers which is
invariant up to scalar multiplication under the action of $\Omega g$.
This combination must be taken between $\psi$'s in  the same sector.
So we consider even $m$ case and odd $m$ case case seperately in the
following.

\begin{itemize}

\item Odd $m$ case:

In this case, $\alpha_J=-\tilde{\alpha}_J\equiv \alpha$ holds.
Therefore the liner combination can be defined as,
\begin{eqnarray}
d^{a}_{r+a/n+\alpha/2n} &\equiv&
\sum_{J=0}^{m-1}D_J\psi^{a,J}_{r+a/n+\alpha/2n},
\nn \\
\tilde{d}^{-a}_{r+a/n+\alpha/2n}&\equiv&(\Omega g\cdot
d)^{a}_{r+a/n+\alpha/2n}.
\end{eqnarray}
If one can find appropriate coefficients $D_J$ such that $d$'s
satisfy,
\begin{equation}
(\Omega g\cdot \tilde{d})^{a}_{r+a/n+\alpha/2n}=\lambda
d^{a}_{r+a/n+\alpha/2n},
\end{equation}
$d^{a}_{r+a/n+\alpha/2n}\tilde{d}^{-a}_{r+a/n+\alpha/2n}$
becomes diagonal under the action of $\Omega g$.
\begin{equation}
\Omega g\cdot(d^{a}_{r+a/n+\alpha/2n}\tilde{d}^{-a}_{r+a/n+\alpha/2n})
=\lambda(d^{a}_{r+a/n+\alpha/2n}\tilde{d}^{-a}_{r+a/n+\alpha/2n}).
\end{equation}
The necessary recursion relation for $D_J$ is
\begin{eqnarray}
\lambda D_{J+2}
&=&{\bf e}\Bigl[
\frac{a+\alpha/2}{n}(p_{J+1}-p_J)
+\frac{\beta_{0,J}}{2}+\frac{\tilde{\beta}_{0,J+1}}{2}
\nn \\ &&
-\sum_{K=0}^{n+p_J-2}\frac{\tilde{\alpha}_{K,J+1}}{2}
-\sum_{K=0}^{n+p_{J+1}-2}\frac{\alpha_{K,J+2}}{2}
\Bigr]D_J.
\end{eqnarray}
By using this recursion $m$ times and from the condition $D_0=D_{2m}$,
the eigenvalue $\lambda_{b}$ for $\Omega g$ can be evaluated as,
\begin{eqnarray}
\lambda_{b}&=&
\be{\frac{\beta^{\{p\}}}{2m}+\frac{b}{m}}, \\
\beta^{\{p\}}&\equiv&
\sum_{l=0}^{m-1}\left(
\beta_{0,J}+\tilde{\beta}_{0,J}
-\sum_{K=0}^{n-2+p_{2l}}\tilde{\alpha}_{K,2l+1}
-\sum_{K=0}^{n-2+p_{2l+1}}\alpha_{K,2l+2}
\right).
\end{eqnarray}

Thus we can evaluate the partition function.
The oscillator part of the partition function is written as,
\begin{equation}
\prod_{r\in \bN-\alpha/2}\left(
1-\be{\frac{\beta^{\{p\}}}{2}}\be{\frac{2m\tau}{n}r}
\right).
\end{equation}
Gathering all factors we obtain the partition function
for Klein bottle as,
\begin{equation}
Z(n,m|\tau)=Z^{-}(2\tau_{n,m,0}).
\end{equation}
This can be interpreted as the partition function for the large Klein
bottle.

\item Even $m$ case:

In this case, we have two different $\alpha$'s corresponding
to even $J$ and odd $J$.
So the eigenvalue for $\Omega g$ is evaluated independently
as odd $m$ case,
Therefore the partition function is evaluated as,
\begin{eqnarray}
&&
\prod_{a=0}^{n-1}\prod_{b=0}^{m/2-1}
Z^{J:{\rm even}}_{a,b}Z^{J:{\rm odd}}_{a,b}
\nn \\
&=&
\prod_{r\in \bN-\alpha_L/2}
\left(
1-\be{\frac{\beta^{\{p\}}_L}{2}}\be{\frac{m\tau -p}{n}r}
\right)
\prod_{r\in \bN-\alpha_R/2}
\left(
1-\be{\frac{\beta^{\{p\}}_R}{2}}\be{\frac{m\tau +p}{n}r}
\right),
\nn \\
&&
\beta^{\{p\}}_L \equiv
\sum_{l=0}^{m/2-1}\Biggl(
\beta_{0,2l}+\beta_{0,2l+1}
-\sum_{K=0}^{n-2+p_{m-2-2l}}\tilde{\alpha}_{K,m-1-2l}
-\sum_{K=0}^{n-2+p_{m-1-2l}}\alpha_{K,m-2l}
\Biggr), \nn \\
&&
\beta_{R}^{\{p\}}\equiv
\sum_{l=0}^{m/2-1}\Biggl(
\beta_{0,2l+1}+\tilde{\beta}_{2l}
-\sum_{K=0}^{n-2-p_{m-1-2l}}\tilde{\alpha}_{K,m-2l}
-\sum_{K=0}^{n-2-p_{m-2-2l}}\alpha_{K,m-2l+1}
\Biggr),
\nn \\
&&p\equiv
\sum_{l=0}^{m/2-1}p_{m-1-2l}-\sum_{l=0}^{m/2-1}p_{m-2l},
\end{eqnarray}
Gathering all factors,
we obtain the partition function for this case,
\begin{equation}
Z(n,m,p|\tau)=Z^{-}(\tau_{n,m,p})Z^{-}(\bar{\tau}_{n,m,p})^{*}.
\end{equation}
This can be interpreted as the partition function for the large torus.

\end{itemize}

\subsection{Generating function of Partition function}

The partition functions for the irreducible sets are calculated for
various topologies of the world-sheets.
We proved that
the interpretation for these partition functions can be made consistent
with the bosonic open string case.
Since the fermionic partition functions are classified in the 
same way as the bosonic case, 
we quote the table in the previous work\cite{Matsuo-Fuji} here.

\begin{center}
\begin{tabular}[b]{|c|c|c|c|c|}\hline
 Short string sector& $n$&$m$ & Long string sector& Partition
function\\\hline
 KB& $*$ & odd &Klein Bottle & $\cZ^{KB}(\tau_{n,m,0})$\\
   & $*$ & even & Torus& $\cZ^{T}(\tau_{n,m,p},\bar\tau_{n,m,p})$\\\hline
 Annulus: $I_A$ & odd & $*$ & Annulus & $\cZ^{A}(\tau_{n,m,0})$ \\
 & even & $*$ & Annulus+M\"obius &
$\cZ^{A}(\tau_{n,m,0})+\cZ^{M}(\tau_{n,m,0})$ \\
 Annulus: $II_A$ & $*$ & $2\times *$ & Klein Bottle &
$\cZ^{KB}(\tau_{n,m/2,0})$\\
 Annulus: $\widetilde{II}_A$ & $*$ & $2\times *$ & Torus&
$\cZ^{T}(\tau_{n,m/2,p},\bar\tau_{n,m/2,p})$\\\hline
 M\"obius: $I_M$ & odd & odd & M\"obius &
$\cZ^{M}(\tau_{n,m,0})$\\
  & odd & even & Annulus& $\cZ^{A}(\tau_{n,m,0})$ \\
  & even & even &Annulus+M\"obius
& $\cZ^{A}(\tau_{n,m,0})+\cZ^{M}(\tau_{n,m,0})$\\
  & even & odd & --- &0\\
 M\"obius: $II_M$ & $*$ & $2\times$odd & Torus &
$\cZ^{T}(\tau_{n,m/2,p^*},\bar\tau_{n,m/2,p^*})$ \\
  & $*$ & $2\times$even & Klein Bottle & $\cZ^{KB}(\tau_{n,m/2,0})$\\
 M\"obius: $\widetilde{II}_M$ & $*$ & $2\times$even & Torus &
$\cZ^{T}(\tau_{n,m/2,p},\bar\tau_{n,m/2,p})$\\
  & $*$ & $2\times$odd & Klein Bottle & $\cZ^{KB}(\tau_{n,m/2,0})$\\\hline
\end{tabular}
\end{center}
In this table, $p$ is an integer from $0$ to $n-1$ and
$p^{*}$ is the half-odd integer from $0$ to $n$.
The modular parameter is defined as $\tau_{n,m,p}\equiv (m\tau-p)/n$,
where $\tau$ is pure imaginary.

The partition functions which appear in this table expressed as,
\begin{eqnarray}
\cZ^{T}(\tau,\bar{\tau}) &=&
({\rm Im}\, \tau)^{-4}\frac{Z^{-}(\tau)}{\eta(\tau)^8}\cdot
\left(\frac{Z^{-}(\bar{\tau})}{\eta(\bar{\tau})^8}\right)^*,
\nn \\
\cZ^{A}(\tau)&=&
N^2(2{\rm Im}\, \tau)^{-4}\frac{Z^{-}(\tau)}{\eta(\tau)^8},
\nn \\
\cZ^{MS}(\tau)&=&
\eta N(2{\rm Im}\, \tau)^{-4}\frac{Z^{-}(\tau+1/2)}{\eta(\tau+1/2)^8},
\nn \\
\cZ^{KB}(\tau)&=& ({\rm Im}\,\tau)^{-4}\frac{Z^{-}(2\tau)}{\eta(2\tau)^8},
\end{eqnarray}
where $Z^{-}$ is defined as (\ref{super partition fcn}) and
we denoted $N$ as the rank of the Chan-Paton group.
When $\eta=1$, this group is $Sp(N)$ and
when $\eta=-1$, this group is $SO(N)$.

By considering the combinatorial factors,
the generating function of the partition function can be calculated as
the closed string case.
The proof for this is same as the bosonic open string case.\cite{Matsuo-Fuji}
These functions can be also interpreted as the DLCQ partition functions.
The result is as follows.
\begin{itemize}
\item Annulus:

When the world-sheet topology is annulus,
the partition function for the $S^N\bR^8$ is defined as,
\begin{equation}
Z_N^{A}(\tau)\equiv \frac{1}{N!}\sum_{h,g\in S_N}
{\rm Tr}_{\cH_{f_1,f_2}}(g\be{\tau L_0}),
\end{equation}
with the constraints, $gf_1 = D_1f_1g$, $gf_2=\cD _2f_2g$,
$f_1^2=f_2^2=\cE$, $h=f_2^{-1}f_1$.
The generating function for this partition function can be written as,
\begin{eqnarray}
&&
\sum_{N=0}^{\infty}\zeta^N Z_N^A(\tau) \nn \\
&=&
{\rm exp}\Biggl(
\sum_{n,m=1}^{\infty}\frac{\zeta^{nm}}{m}\cZ^A(\tau_{n,m,0})
+\sum_{n,m=1}^{\infty}\frac{\zeta^{2nm}}{m}\cZ^{MS}(\tau_{2n,m,0})\nn \\
&&
+\sum_{n,m=1}^{\infty}\frac{\zeta^{2nm}}{2m}\cZ^{KB}(\tau_{n,m,0})
+\sum_{n,m=1}^{\infty}\frac{\zeta^{2nm}}{2nm}\sum_{p=0}^{n-1}
\cZ^{T}(\tau_{n,m,p},\bar{\tau}_{n,m,p})
\Biggr).
\end{eqnarray}

\item M\"obius strip:

When the world-sheet topology is M\"obius strip,
the partition function for the $S^N\bR^8$ is defined as,
\begin{equation}
Z_N^{MS}(\tau)\equiv \frac{1}{N!}\sum_{h,g\in S_N}
{\rm Tr}_{f_1,f_2}(\Omega g\be{\tau L_0}),
\end{equation}
with the constraints,
$f_2=\cC gf_1^{-1}g^{-1}$, $h=gf_1g^{-1}f$, $f_1^2=\cE$,
$\left[\cC,g\right]=0$, $\left[\cC g^2,f_1\right]= 0$.
The generating function for this partition function can be written as,
\begin{eqnarray}
&&
\sum_{N=0}^{\infty}\zeta^N Z_N^{MS}(\tau) \nn \\
&=&
{\rm exp}\Biggl(
\sum_{n,m=1}^{\infty}\frac{\zeta^{2nm}}{2m}\cZ^A(\tau_{n,2m,0})
\nn \\
&&
+\sum_{n,m=1}^{\infty}\frac{\zeta^{(2n-1)m}}{m}\cZ^{MS}(\tau_{2n-1,m,0})
+\sum_{n,m=1}^{\infty}\frac{\zeta^{2nm}}{2m}\cZ^{KB}(\tau_{n,m,0})
\nn \\
&&
+\sum_{n,m=1}^{\infty}\frac{\zeta^{2nm}}{2nm}\sum_{p=0}^{n-1}
\cZ^{T}(\tau_{n,m,p+n/2},\bar{\tau}_{n,m,p+n/2})
\Biggr).
\end{eqnarray}

\item Klein bottle:

When the world-sheet topology is Klein bottle,
the partition function for the $S^N\bR^8$ is defined as,
\begin{equation}
Z_N^{MS}(\tau)\equiv \frac{1}{N!}\sum_{h,g\in S_N}
{\rm Tr}_h(\Omega g\be{\tau L_0}),
\end{equation}
with a constraint $hg=gh^{-1}$.
The generating function for this partition function can be written as,
\begin{eqnarray}
&&
\sum_{N=0}^{\infty}\zeta^N Z_N^{KB}(\tau) \nn \\
&=&
{\rm exp}\Biggl(
\sum_{n,m=1}^{\infty}\frac{\zeta^{n(2m-1)}}{2m-1}\cZ^{KB}(\tau_{n,2m-1,0})
\nn \\ &&
+\sum_{n,m=1}^{\infty}\frac{\zeta^{2nm}}{2nm}\sum_{p=0}^{n-1}
\cZ^{T}(\tau_{n,2m,p},\bar{\tau}_{n,2m,p})
\Biggr).
\end{eqnarray}

\end{itemize}

\section{Boundary State for Fermionic Fields}

In the previous section, we considered the open string on
the permutation orbifold in the open string sector.
In this section, we will construct the boundary state and see how the
change of the world-sheet topology occurs.
Here we concentrate on the case of $S^{N}\bR^1$,
since the generalization to $S^{N}\bR^8$ is clear.

\subsection{Boundary state for the irreducible sets}

In terms of the closed string sector,
we will consider the irreducible combination of the twist $g$ along the
space direction and the boundary twist $f$.
The closed string Hilbert space is specfied by the twist along the
space direction.
As discussed in section 2, we can decompose the Hilbert space $\cH_g$
into that of the conjugacy class and from the condition $fg=\cD gf$,
the boundary twist belongs to its centralizer group.
Thus the irreducible combination of $(g,f)$ can be written as,
\begin{equation}
g=\diag(A_0\cyclb_{\tn},\cdots ,A_{\tm-1}\cyclb_{\tn}),
\quad
f=\diag(F_0\cyclb_{\tn}^{\tp_0},\cdots
,F_{\tm-1}\cyclb_{\tn}^{\tp_{m-1}})\cdot\cycl_{\tm}.
\label{irr combi}
\end{equation}

For this irreducible combination of the twists, further constraint
(\ref{LR symmetry F}) must be imposed.
This constraint needs the value of $\tm$ to be $1$ or $2$.
As a result, we can classify the irreducible sets into three types.
\begin{itemize}
\item $\tm=1$, $\tn$:odd case.

In this case, the boundary twist $f$ can be written
as $f=F_0\cyclb_0^{\tp_0}$.
The condition $f^2=\cE$ imposes the condition,
\begin{equation}
2\tp_0\equiv 0, \quad \mod{\tn}.
\label{2p0}
\end{equation}
For odd $\tm$ case, we have only one solution $\tp_0\equiv 0$ (mod $\tn$).

\item $\tm=1$, $\tn$:even case.

In this case, the boundary twist $f$ is same as above case.
But there are two solutions for (\ref{2p0}) as
$\tp_0\equiv 0$ (mod $\tn$) and $\tp_0\equiv \tn/2$ (mod $\tn$).

\item $\tm=2$ case.

In this case, the boundary twist $f$ is written as
$f=\diag(F_0\cyclb_{\tn},F_1\cyclb_{\tn})\cdot\cycl_{2}$.
The condition $f^2=\cE$ imposes the condition,
\begin{equation}
\tp_0+\tp_1\equiv 0, \quad \mod{\tn}.
\end{equation}

\end{itemize}

We will construct the boundary states corresponding to these boundary
twists.
Though we imposed the condition $fg=\cD gf$ in the open string setor,
we impose the condition $fg=gf$ here
and the non-commuting condition is imposed on another boundary.

\vskip 2mm
(i)\underline{{\it Boundary states for the long string}}:

In this case, the combination of twists is $(g,f)=(A\cyclb_{\tn},F)$.
The consistency condition $fg=gf$ implies,
\begin{equation}
\phi\equiv \phi_{\tI}, \quad f=\be{\frac{\phi}{2}}\cdot {\bf 1}_{\tn}.
\end{equation}
If we use the mode expansion for the closed string (\ref{mode T2}),
the boundary state can be written as,
\begin{equation}
\Bket{A\cyclb_{\tn}}{F}
=\exp\left(
-i\be{\frac{\phi}{2}}\sum_{r=1}^{\infty}\sum_{a=0}^{\tn-1}
\psi^{(a)}_{-r+a/\tn +\alpha/2\tn}
\tilde{\psi}^{(\tn-a)}_{-r+a/\tn +\alpha/2\tn}
\right)|0\rangle_{\tn},
\end{equation}
where we denote $\alpha=\sum_{\tI=0}^{\tn-1}\alpha_{\tI}.$

We introduce the long string oscillator of length $\tn$ as,
\begin{equation}
\Psi_{\tn r+a+\alpha/2}\equiv \psi^{(a)}_{r+a/\tn +\alpha/2\tn},
\quad
\tilde{\Psi}_{\tn r+a+\alpha/2}\equiv
\tilde{\psi}^{(\tn-a)}_{r+a/\tn +\alpha/2\tn}.
\end{equation}
These oscillators satisfy the commutation relation
$\{\Psi_r,\Psi_s\}=\delta_{r+s,0}$.
The commutation relation with Hamiltonian is modified to
$[L_0,\Psi_r]=-\frac{r}{\tn}\Psi_r$.
In terms of these variables, the boundary state can be rewritten as,
\begin{equation}
|B\rangle_{\tn}\equiv \exp\left(
-i\be{\frac{\phi}{2}}\sum_{r\in \bN-\alpha/2}\Psi_{-r}
\tilde{\Psi}_{-r}
\right)|0\rangle_{\tn}.
\end{equation}
This boundary state can be interpreted as the boundary state
for the long string.

\vskip 2mm
(ii)\underline{{\it Cross-cap states for the long string}}:

If $\tn$ is even, there is a combination of twists
$(g,f)=(A\cyclb_{\tn},F\cyclb_{\tn}^{\tn/2})$.
The consistency condition $fg=gf$ implies,
\begin{equation}
\phi_{\tI}-\phi_{\tI+1}=\alpha_{\tI}-\alpha_{\tI+\tl},
\label{fg cond BC}
\end{equation}
where we denoted as $\tl\equiv\tn/2$.
Under this relation, the action of $f$
on the diagonalized field $\psi^{(a)}$ becomes,
\begin{eqnarray}
f\cdot\psi^{(a)}
&=&\be{\frac{\beta}{2}}(-1)^{a+\alpha/2}\psi^{(a)}, \\
\beta&\equiv&\phi_{0}-\sum_{\tK=0}^{\tI-1}\alpha_{\tK}.
\end{eqnarray}
In deriving above action, we used the relation,
\[
\phi_{\tI}-\sum_{\tK=0}^{\tI+\tl-1}\alpha_{\tK}
+\sum_{\tK=0}^{\tI-1}\alpha_{\tK}
=\phi_0-\sum_{\tK=0}^{\tl-1}\alpha_{\tK},
\]
which is derived from the relation (\ref{fg cond BC}).

The boundary state for this twist can be written as,
\begin{eqnarray}
&&
\Bket{A\cyclb_{\tn}}{F\cyclb_{\tn}^{\tn/2}}\equiv
|C\rangle_{\tn}, \nn \\
&=&
\exp\left(
-i\be{\frac{\beta}{2}}\sum_{r=1}^{\infty}\sum_{a=0}^{\tn-1}(-1)^{a+\alpha/2}
\psi^{(a)}_{-r+a/\tn+\alpha/2\tn}
\tilde{\psi}^{(\tn-a)}_{-r+a/\tn+\alpha/2\tn}
\right)|0\rangle_{\tn}, \nn \\
&=&
\exp\left(
-i\be{\frac{\beta}{2}}\sum_{r\in\bN-\alpha/2}
(-1)^{r}\Psi_{-r}\tilde{\Psi}_{-r}
\right)|0\rangle_{\tn}.
\end{eqnarray}
This boundary state can be intepreted as the cross-cap state for the
long string.
This is the origin of the topology change
from the oriented world-sheet to the unoriented one.

\vskip 2mm
(iii)\underline{{\it Joint state}}:

In this case, $\tm=2$.
and the combination of the twists is,
\begin{equation}
g=\left(
\begin{array}{cc}
A_0\cyclb_{\tn} & 0\\
0&A_1\cyclb_{\tn}
\end{array}
\right),
\quad
f=\left(
\begin{array}{cc}
0 & F_0\cyclb_{\tn}^{\tp_0} \\
F_1\cyclb_{\tn}^{\tp_1} & 0
\end{array}
\right).
\end{equation}
The consistency condition $fg=gf$ implies,
\begin{eqnarray}
&&\phi_{\tI,\tJ}-\phi_{\tI+1,\tJ}=
\alpha_{\tI,\tJ}
-\alpha_{\tI+\tp_{\tJ},\tJ+1},
\label{fg cond BJ}\\
&& \tI=0,\cdots ,\tn-1, \quad
\tJ=0,1. \nn
\end{eqnarray}
Summing up $\tI$, we obtain $\alpha\equiv \alpha_{\tJ}$.

Under this relation, the acion of $f$ on the diagonalid field
$\psi^{(a,\tJ)}$ can be written as,
\begin{eqnarray}
f\cdot\psi^{(a,\tJ)}&=&
\be{\frac{\beta^{\tp}}{2}+\frac{(a+\alpha/2)\tp_{\tJ}}{\tn}}
\psi^{(a,\tJ+1)}, \\
\beta^{\tp}_{\tJ}&\equiv&
\phi_{0,\tJ}-\sum_{\tK=0}^{\tp_{\tJ}-1}\alpha_{\tK,\tJ+1}.
\end{eqnarray}
In deriving above action, we used the relation,
\[
\phi_{\tI,\tJ}-\sum_{\tK=0}^{\tI+\tp_{\tJ}-1}\alpha_{\tK,\tJ+1}
+\sum_{\tK=0}^{\tI-1}\alpha_{\tK,\tJ}
=\phi_{0,\tJ}-\sum_{\tK=0}^{\tp_{\tJ}-1}\alpha_{\tK,\tJ+1},
\]
which is derived
from the relation (\ref{fg cond BJ}).

The boundary state for this twist can be written as,
\begin{eqnarray}
&&
\Bket{g}{f}\equiv |J(12),\tp\rangle_{\tn}, \nn \\
&=&
\exp\Biggl(
i\be{\frac{\beta^{\tp}_0}{2}}\sum_{r=1}^{\infty}\sum_{a=0}^{\tn-1}
\be{\frac{(a+\alpha/2)\tp}{\tn}}
\psi^{(a,1)}_{-r+a/\tn+\alpha/2\tn}
\tilde{\psi}^{(\tn-a,2)}_{-r+a/\tn+\alpha/2\tn}
\nn \\
&&
+i\be{\frac{\beta^{\tp}_1}{2}}\sum_{r=1}^{\infty}\sum_{a=0}^{\tn-1}
\be{\frac{-(a+\alpha/2)\tp}{\tn}}
\psi^{(a,2)}_{-r+a/\tn+\alpha/2\tn}
\tilde{\psi}^{(\tn-a,1)}_{-r+a/\tn+\alpha/2\tn}
\Biggr)|0\rangle_{\tn}, \nn \\
&=&
\exp\Biggl(
i\be{\frac{\beta^{\tp}_0}{2}}\sum_{r\in\bN-\alpha/2}\be{\frac{r\tp}{\tn}}
\Psi^{(1)}_{-r}\tilde{\Psi}^{(2)}_{-r} \nn \\
&&
+i\be{\frac{\beta^{\tp}_1}{2}}\sum_{r\in\bN-\alpha/2}\be{-\frac{r\tp}{\tn}}
\Psi^{(2)}_{-r}\tilde{\Psi}^{(1)}_{-r}
\Biggr)|0\rangle_{\tn},
\end{eqnarray}
where we denoted $\tp\equiv\tp_0 =-\tp_1$ and
$\Psi^{(\tJ)}$ ($\tJ$=1,2) is the oscillator for two long stings.
This boundary state represents the interconnection of two long strings
at the boundary.
So we call this state as ``Joint state''.
\footnote{
By construction, the joint state represents the single type of the
connected string.
So, if we take the inner product between the products of only joint states,
the amplitude becomes the left-right symmetric (Type 0 string)
character on the torus.
This is originated from $[f,g]=0$.
Since the joint state is the local representation for the
interconnection of the two strings, the global representation needs
the generalization of the consistency condition.
Therefore,
in order to realize Type I superstring theory,
we need the $\bZ_2$-extended boundary conditions (\ref{fg cond A}).
}

\subsection{Cross-cap state for the irreducible set}

As considered in the open string analysis,
the cross-cap state satisfies $g=f^2\cC$.
The irreducible combination of the twists
can be found from (\ref{irr combi}).
The cross-cap condition impose the value of $\tm$ to be $1$ or $2$.
The boundary twists are classified as follows.
\begin{itemize}
\item $\tm=1$ case.

In this case, the boundary twists $f$ can be written as
$f=F_0\cyclb_{\tn}^{\tp_0}$.
The cross-cap condition $g=f^2\cC$ imposes the following condition,
\begin{equation}
2\tp_0\equiv 1, \quad \mod{\tn}.
\end{equation}
For odd $\tn$ case, we have one solution $\tp_{0}=(\tn-1)/2$.
For even $\tn$ case, we have no solution for above condition.

\item $\tm=2$ case.

In this case, the boundary twist $f$ is written as
$f=\diag(F_0\cyclb_{\tn},F_1\cyclb_{\tn})\cdot\cycl_{2}$.
The cross-cap condition $g=f^2\cC$ imposes the following condition,
\begin{equation}
\tp_0+\tp_1\equiv 1, \quad \mod{\tn}.
\end{equation}
We will construct the cross-cap states corresponding to these boundary
twists.

\end{itemize}

\vskip 2mm
(i)\underline{{\it Cross-cap state for long string}}:

When $\tm=1$, there is a combination of twists
$(g,f)=(A\cyclb_{\tn},F\cyclb_{\tn}^{(\tn+1)/2})$
for odd $\tn$.
The constraint $[f,\cC]=0$ and $[f,g]=0$ implies
$\cC=\diag\left(\be{\frac{c}{2}},\cdots ,\be{\frac{c}{2}}
\right)$.

From the constraint $g=f^2\cC$, we obtain the condition as,
\begin{equation}
\alpha_{\tI}=c+\phi_{\tI}+\phi_{\tI+(\tn+1)/2}.
\end{equation}
The action of $f$ on the diagonalized field $\psi^{(a)}$ is
\[
(f\cdot\psi)^{(a)}
=\be{\frac{a+\alpha/2}{\tn}\cdot\frac{\tn+1}{2}
+\frac{(\tn+1)c}{4}
+\frac{\phi}{2}}\psi^{(a)}.
\]

The cross-cap condition is written as,
\begin{equation}
\left(\psi^{(a)}(\sigma^0,\sigma^1+\pi)
+if\cdot\tilde{\psi}^{(\tn-a)}(\sigma^0,\sigma^1)\right)
\Ckets{g}{f}=0.
\label{cc cond}
\end{equation}
This condition is solved as,
\begin{eqnarray}
\Ckets{g}{f}&=&
\exp\Biggl(
-i\be{\frac{\phi}{2}+\frac{(\tn+1)c}{4}}\sum_{s=1}^{\infty}\sum_{a=0}^{\tn-1
}
\nn \\
&&\times
\be{\frac{s-a-\alpha/2}{2}}
\psi^{(a)}_{-s+a/\tn+\alpha/2\tn}
\tilde{\psi}^{(\tn-a)}_{-s+a/\tn+\alpha/2\tn}
\Biggr)|0\rangle_{\tn},
\nn \\
&=&
\exp\left(
-i\be{\frac{\phi}{2}+\frac{(\tn+1)c}{4}}
\sum_{r\in\bN-\alpha/2}(-1)^{r}\Psi_{-r}\tilde{\Psi}_r
\right)
|0\rangle_{\tn}.
\end{eqnarray}
This can be interpreted as the cross-cap state for the long string.

\vskip 2mm
(ii)\underline{{\it Joint state}}:

For $\tm=2$, we have the twists as,
\begin{equation}
g=\left(
\begin{array}{cc}
A_0\cyclb_{\tn} & 0\\
0&A_1\cyclb_{\tn}
\end{array}
\right),
\quad
f=\left(
\begin{array}{cc}
0 & F_0\cyclb_{\tn}^{\tp_0} \\
F_1\cyclb_{\tn}^{\tp_1} & 0
\end{array}
\right).
\end{equation}
The condition $[f,\cC]=0$ and $[g,f]=0$ implies
$\cC=\diag(C,C)$, $C=\diag(\be{\frac{c}{2}},\cdots,\be{\frac{c}{2}})$.

From the constraint $g=f^2\cC$, we obtain the condition as,
\begin{equation}
\alpha_{\tI,\tJ}=c+\phi_{\tI,\tJ}+\phi_{\tI+\tp_{\tJ},\tJ+1}.
\end{equation}
Summing up $\tI$, we obtain the relation $\alpha\equiv\alpha_{\tJ}$.

The action of $f$ on the diagonalized field $\psi^{(a,\tJ)}$ is
written as,
\begin{eqnarray}
(f\cdot \psi)^{(a,\tJ)}
&=&
\be{\frac{(a+\alpha/2)}{\tn}+\frac{\beta^{\tp}_{\tJ}}{2}}\psi^{(a,\tJ+1)},
\\
\beta^{\tp}_{\tJ}&\equiv&
\phi_{0,\tJ}-\sum_{\tK=0}^{\tp_{\tJ}-1}\alpha_{\tK,\tJ+1}.
\end{eqnarray}

The cross-cap condition (\ref{cc cond}) is solved as,
\begin{eqnarray}
&&
\Ckets{g}{f} \nn \\
&=&
\exp\Biggl(
i\be{\frac{\beta^{\tp}_0}{2}}\sum_{s=1}^{\infty}\sum_{a=0}^{\tn-1}
(-1)^{-s+a/\tn+\alpha/2\tn}\be{(\frac{a}{\tn}+\frac{\alpha}{2\tn})\tp}
\psi^{(a,0)}_{-s+a/\tn+\alpha/2\tn}
\tilde{\psi}^{(a,1)}_{-s+a/\tn+\alpha/2\tn} \nn \\
&&
+i\be{\frac{\beta^{\tp}_1}{2}}\sum_{s=1}^{\infty}\sum_{a=0}^{\tn-1}
(-1)^{-s+a/\tn+\alpha/2\tn}\be{-(\frac{a}{\tn}+\frac{\alpha}{2\tn})(\tp+1)}
\psi^{(a,1)}_{-s+a/\tn+\alpha/2\tn}
\tilde{\psi}^{(a,0)}_{-s+a/\tn+\alpha/2\tn}
\Biggr)|0\rangle_{\tn}, \nn \\
&=&
\exp\Biggl(
i\be{\frac{\beta^{\tp}_0}{2}}\sum_{s=1}^{\infty}\sum_{a=0}^{\tn-1}
(-1)^{-s}\be{(\frac{a}{\tn}+\frac{\alpha}{2\tn})(\tp+1/2)}
\Psi^{(a,1)}_{-\tn s+a+\alpha/2}
\tilde{\Psi}^{(a,2)}_{-\tn s+a+\alpha/2} \nn \\
&&
+i\be{\frac{\beta^{\tp}_1}{2}}\sum_{s=1}^{\infty}\sum_{a=0}^{\tn-1}
(-1)^{-s}\be{-(\frac{a}{\tn}+\frac{\alpha}{2\tn})(\tp+1/2)}
\Psi^{(a,1)}_{-\tn s+a+\alpha/2}
\tilde{\Psi}^{(a,2)}_{-\tn s+a+\alpha/2}
\Biggr)|0\rangle_{\tn}.
\end{eqnarray}
Two short strings are connedted at this boundary in curious way.
The factor $\tp+1/2$ is the origin of the ``half-twist'' in the
large torus amplitude.

\subsection{Inner product between boundary states}

In order to reproduce the open string amplitudes which we calculated
in the previous section, we will consider the inner products
between the boundary states.
If the irriducibility condition for the periodicity and boundary twist
is loosened,
the open string partition functions for the irreducible sets can be
realized.
So we use the following general twists for the boundary conditions,
\begin{eqnarray}
f_1=\diag(F^{(1)}_0\cyclb_{\tn}^{\tp^{(1)}_0},\cdots ,
F^{(1)}_{\tm-1}\cyclb_{\tn}^{\tp^{(1)}_{\tm-1}})
\cdot\cycl_{\tm}^{\tq_1}\inv_{\tm}, \nn \\
f_2=\diag(F^{(2)}_0\cyclb_{\tn}^{\tp^{(2)}_0},\cdots
,F^{(2)}_{\tm-1}\cyclb_{\tn}^{\tp^{(2)}_{\tm-1}})
\cdot\cycl_{\tm}^{\tq_2}\inv_{\tm},
\label{bdy f}
\end{eqnarray}
where these twists satisfy the condition $f_i^2=\cE$.

For the consistency of the amplitude,
the condition $[h,g]=0$ must be satisfied.
This condition gives the following constraint for $\alpha$ and $\phi^{(i)}$,
\begin{eqnarray}
&&
\alpha_{\tI,\tJ}-\alpha_{\tI+\tp^{(1)}_{\tJ}
+\tp^{(2)}_{\tm-1-\tJ+\tq_1},\tJ\tq_2-\tq_1}
\nn \\
&=&
\phi^{(1)}_{\tI,\tJ}-\phi^{(1)}_{\tI+1,\tJ}
+\phi^{(2)}_{\tI+\tp^{(1)}_{\tJ},\tm-1-\tJ+\tq_1}
-\phi^{(2)}_{\tI+\tp^{(1)}_{\tJ}+1,\tm-1-\tJ+\tq_2},
\label{ab cond bdy}
\end{eqnarray}
where we defined $h\equiv f_2f_1$.
On the other hand, the condition $f_i^2=\cE$ gives the following
constraint for $p_{\tJ}$ and $q_{\tJ}$ as,
\begin{equation}
\tp^{(i)}_{\tJ}+\tp^{(i)}_{\tm-1-\tJ+\tq_{i}}\equiv 0 \quad \mod{\tn}.
\label{p cond}
\end{equation}

To classify the amplitudes, we need to interpret the boundary condition
physically.
If there are string bits which is invariant under the boundary
twist (\ref {bdy f}), the long string has the loose ends.
The condition which the long string has the loose ends is,
\begin{equation}
\tl\equiv \tm-1-\tl+\tq_{i} \quad \mod{\tn}.
\label{loose end}
\end{equation}
From the solution for this condition, we can classify the
boundary twists into two types.
\begin{itemize}

\item {\it Long open string boundary}:

If the boundary twists $f_1$ and $f_2$ admit two loose ends which
satisfy the condition (\ref{loose end}), this sector expresses the
long open string.
The condition for this case is that $\tm$ is even and $\tq_1-\tq_2$ is
odd or $\tm$ is odd.
So we set $\tq_1=1$ and $\tq_2=0$ as the representative.

Before evaluating the eigenvalue of $h=f_2f_1$,
the constraint (\ref{ab cond bdy}) implies
$\alpha\equiv \alpha_{\tJ}$.
So the eigenvalue for $f_2f_1$ can be calculated as before,
\begin{eqnarray}
\mu_{a,b}&=&
\be{\frac{(a+\alpha/2)\tp}{\tn\tm}+
\frac{\beta^{\{\tp^{(1)},\tp^{(2)}\}}}{2\tm}+\frac{b}{\tm}}
, \nn \\
\tp &\equiv&
\sum_{\tJ=0}^{\tm-1}
(\tp^{(1)}_{\tJ}+\tp^{(2)}_{\tJ}), \nn \\
\beta^{\{\tp^{(1)},\tp^{(2)}\}}
&\equiv &
\sum_{\tJ=0}^{\tm-1}\left(\phi^{(1)}_{0,\tJ}
+\phi^{(2)}_{\tp^{(1)}_{\tJ},\tm-\tJ}
-\sum_{\tK=0}^{\tp^{(1)}+\tp^{(2)}_{\tm-\tJ-1}-1}
\alpha_{\tK,\tJ-1}
\right).
\end{eqnarray}
Because of (\ref{p cond}), $\tp$ depends only on the value
at the loose ends $\tp=\tp^{(1)}_{\tl}+\tp^{(2)}_{\tl^{\prime}}$.
For annulus and Klein bottle, we have $\tp=0$ (mod $\tn$).
For M\"obius strip, we have $\tp=\tn/2$ (mod $\tn$).

In the explicit evaluation, we use the following formulae,
\begin{equation}
\langle0|e^{-if_2^{I,J}\psi_I\tilde{\psi}_J q}
e^{-if_1^{K,L}\psi_K^{\dagger}\tilde{\psi}_L^{\dagger}}
|0\rangle
=\prod_{a}(1+\mu_a q),
\end{equation}
where oscillators satisfy the commutation relations
$\{\psi_I,\psi_J^{\dagger}\}=
\{\tilde{\psi}_I,\tilde{\psi}_J^{\dagger}\}
=\delta_{I,J}$ and $\mu_a$ is the eigenvalue for $f_2f_1$.

Thus the oscillator contributions of annulus and Klein bottle amplitudes
are evaluated as,
\begin{eqnarray}
&&
\Bbra{g}{f_2}q^{L_0+\tilde{L}_0}\Bket{g}{f_1}
\nn \\
&=&
\prod_{s=1}^{\infty}\prod_{a=0}^{\tn-1}\prod_{b=0}^{\tm-1}
\left(
1+\be{\frac{\beta^{\{\tp^{(1)},\tp^{(2)}\}}}{2\tm}}
\be{\frac{b}{\tm}}
q^{2(s-a/\tn-\alpha/2\tn)}
\right), \nn \\
&=&
\prod_{r \in \bN-\alpha/2}\left(
1-\be{\frac{\beta^{\{\tp^{(1)},\tp^{(2)}\}}+\tm}{2}}q^{2\tm r/\tn}
\right).
\end{eqnarray}
For M\"obius strip, the oscillator contribution is written as,
\begin{eqnarray}
&&
\Bbra{g}{f_2}q^{L_0+\tilde{L}_0}\Bket{g}{f_1}
\nn \\
&=&
\prod_{s=1}^{\infty}\prod_{a=0}^{\tn-1}\prod_{b=0}^{\tm-1}
\left(
1+\be{\frac{\beta^{\{\tp^{(1)},\tp^{(2)}\}}}{2\tm}}
\be{\frac{b}{\tm}+\frac{a+\alpha/2}{2\tm}}
q^{2(s-a/\tn-\alpha/2\tn)}
\right), \nn \\
&=&
\prod_{r \in\bN-\alpha/2}\left(
1-
(-1)^{r}\be{\frac{\beta^{\{\tp^{(1)},\tp^{(2)}\}}+\tm}{2}}q^{2\tm r/\tn}
\right).
\end{eqnarray}

Thus we have reproduced the annulus, M\"obius strip and Klein bottle
amplitudes for the long string from the inner products of the
boundary states.

\item {\it Long closed string boundary}:

If the boundary twists $f_1$ and $f_2$ admit no loose end,
this sector expresses the long closed string.
In order not to have any solutions for the condition (\ref{loose end}),
the necessary condition is that $\tm$ is even and both $\tq_1$ and $\tq_2$
is even.
So we set $\tm$ is even and $\tq_1=2$ and $\tq_2=0$ as the representative.

From the consistency conditions $f^2=\cE$ and $[h,g]=0$,
we have the consistency condition for
the boundary twists for the torus amplitude as,
\begin{eqnarray}
&&
\tp^{(1)}_{\tJ}+\tp^{(1)}_{\tm-1-\tJ+2}\equiv 0, \quad
\tp^{(2)}_{\tJ}+\tp^{(2)}_{\tm-1-\tJ}\equiv 0,  \\
\label{p cond bt}
&&
\alpha_{\tI,\tJ}
-\alpha_{\tI+\tp^{(1)}_{\tJ}+\tp^{(2)}_{\tm-1-\tJ+2},\tJ-2}
\nn \\
&=&
\phi^{(1)}_{\tI,\tJ}
-\phi^{(1)}_{\tI+1,\tJ}
+\phi^{(2)}_{\tI+\tp^{(1)}_{\tJ},\tm-1-\tJ+2}
-\phi^{(2)}_{\tI+\tp^{(1)}_{\tJ}+1,\tm-1-\tJ}.
\label{ab cond bt}
\end{eqnarray}
Summing up $\tI$ in (\ref{ab cond bt}), we have
$\alpha_{\tJ}=\alpha_{\tJ-2}$.
Therefore the eigenvalue for $f_2f_1$ should be calssified by $\tJ$.

For even $\tJ$, the eigenvalue can be calculated as,
\begin{eqnarray}
\mu^{L}_{a,b}&=&
\be{\frac{a+\alpha_L/2}{\tn}\cdot \frac{\tp_L}{\tk}
+\frac{\beta_L^{\{\tp^{(1)},\tp^{(2)}\}}}{2\tk}
+\frac{b}{\tk}}, \nn \\
\beta_L^{\{\tp^{(1)},\tp^{(2)}\}}&\equiv &
\sum_{\tJ=0}^{\tk-1}\left(
\phi^{(1)}_{0,2\tJ}
+\phi^{(2)}_{\tp^{(1)}_{2\tJ},\tm-2\tJ+1}
-\sum_{\tK=0}^{\tp^{(1)}_{2\tJ}+\tp^{(2)}_{\tm-2\tJ+1}-1}
\alpha_{\tK,2\tJ-2}
\right), \nn \\
\alpha_L&\equiv& \alpha_{2\tJ}, \nn \\
\tp_L&\equiv&
\sum_{\tJ=0}^{\tk-1}(\tp^{(1)}_{2\tJ}+\tp^{(2)}_{\tm-2\tJ+1}),
\end{eqnarray}
where we denote $\tk\equiv \tm/2$.

For odd $\tJ$, the eigenvalue can be calculated as,
\begin{eqnarray}
\mu^{R}_{a,b}&=&
\be{\frac{a+\alpha_R/2}{\tn}\cdot \frac{\tp_R}{\tk}
+\frac{\beta_R^{\{\tp^{(1)},\tp^{(2)}\}}}{2\tk}
+\frac{b}{\tk}}, \nn \\
\beta_R^{\{\tp^{(1)},\tp^{(2)}\}}&\equiv &
\sum_{\tJ=0}^{\tk-1}\left(
\phi^{(1)}_{0,2\tJ+1}
+\phi^{(2)}_{\tp^{(1)}_{2\tJ+1},\tm-2\tJ}
-\sum_{\tK=0}^{\tp^{(1)}_{2\tJ+1}+\tp^{(2)}_{\tm-2\tJ}-1}
\alpha_{\tK,2\tJ-1}
\right), \nn \\
\alpha_R&\equiv& \alpha_{2\tJ+1}, \nn \\
\tp_R&\equiv&
\sum_{\tJ=0}^{\tk-1}(\tp^{(1)}_{2\tJ+1}+\tp^{(2)}_{\tm-2\tJ}).
\end{eqnarray}
From the consistency condition (\ref{p cond bt}), we have
$\tp_L=-\tp_R\equiv \tp$.

The oscillator contribution for the torus amplitude is written as,
\begin{eqnarray}
&&
\Bbra{g}{f_2}q^{L_0+\tilde{L}_0}\Bket{g}{f_1}
\nn \\
&=&
\prod_{s=1}^{\infty}\prod_{a=0}^{\tn-1}\prod_{b=0}^{\tk-1}
\left(
1+\be{\frac{\beta_L^{\{\tp^{(1)},\tp^{(2)}\}}}{2\tk}}
\be{\frac{b}{\tk}+\frac{(a+\alpha_L/2)\tp}{\tk\tn}}
q^{2(s-a/\tn-\alpha_L/2\tn)}
\right) \nn \\
&&
\times \left(
1+\be{\frac{\beta_R^{\{\tp^{(1)},\tp^{(2)}\}}}{2\tk}}
\be{\frac{b}{\tk}-\frac{(a+\alpha_R/2)\tp}{\tk\tn}}
q^{2(s-a/\tn-\alpha_R/2\tn)}
\right) \nn \\
&=&
\prod_{r\in \bN-\alpha_L/2}\left(
1-\be{\frac{\beta_L^{\{\tp^{(1)},\tp^{(2)}\}}+\tk}{2}}
\be{\frac{2\tk\tau-\tp}{\tn}r}
\right) \nn \\
&& \times
\prod_{r \in\bN-\alpha_R/2}\left(
1-\be{\frac{\beta_R^{\{\tp^{(1)},\tp^{(2)}\}}+\tk}{2}}
\be{\frac{2\tk\tau+\tp}{\tn}r}
\right).
\end{eqnarray}
Thus we reproduced the torus amplitude from the inner product of the
boundary states.

\end{itemize}

\section{Tadpole Condition}

In usual string theory, the consistent theory has no the divergence
which comes from the dilaton tadpole.
Here we consider the tadpole cancellation condition and
determine the consistent Chan-Paton gauge group for
the superstring theory on the permutation orbifold.

The usual GSO-projected boundary and cross-cap states
for NS-$\widetilde{{\rm NS}}$ sector is written as,\footnote{
The boundary/cross-cap states expressed in this section
is the direct product of the bosonic boundary/cross-cap states
which we constructed in the previous paper\cite{Matsuo-Fuji}
and the bosonic boundary/cross-cap states which constructed in the
previous section for $S^N\bR^8$.
}
\begin{eqnarray}
|\,B\rangle\!\rangle^{NS}
&=&
|\,B:+\rangle\!\rangle^{NS}-|\,B:-\rangle\!\rangle^{NS},
\nn \\
|\,C \rangle\!\rangle^{NS}
&=&
|\,C:+\rangle\!\rangle^{NS}-|\,C:-\rangle\!\rangle^{NS},
\end{eqnarray}
where we denoted as
$|\,B(C):\pm\rangle\!\rangle^{NS}\equiv
|\,B(C):\alpha=1,f=\pm 1\rangle\!\rangle$.
The tadpole cancellation condition for these states is
\begin{equation}
|\,B\rangle\!\rangle_{0}^{NS} -i |\,C\rangle\!\rangle_{0}^{NS} =0,
\end{equation}
where we denoted subscript $0$ for the restriction to the massless part.

In our case, in order to construct the GSO-projected boundary states,
we will consider the GSO-projected boundary, cross-cap and joint
states and combine them.
To accomplish the tadpole cancellation between the boundary states
which represents the boundary and cross-cap state,
we restrict $\tn$ to be even.
The action of $(-1)^{F}$ on these states can be written as,
\begin{eqnarray}
&&
(-1)^{F}|\,B(C):,\eta\rangle\!\rangle_{\tn}^{NS}
=-|\,B(C):-\eta\rangle\!\rangle_{\tn}^{NS}, \nn \\
&&
(-1)^{F_1}|\,J:(\eta_1,\eta_2),\tp\rangle\!\rangle_{\tn}^{NS}
=-|\,J:(-\eta_1,\eta_2),\tp\rangle\!\rangle_{\tn}^{NS}, \nn \\
&&
(-1)^{F_2}|\,J:(\eta_1,\eta_2),\tp\rangle\!\rangle_{\tn}^{NS}
=-|\,J:(\eta_1,-\eta_2),\tp\rangle\!\rangle_{\tn}^{NS}. \nn
\end{eqnarray}
The action of $(-1)^{\tilde{F}}$ is similar as above.
Therefore the GSO-projected boundary states for the irreducible combination
is,
\begin{eqnarray}
|\,B\rangle\!\rangle_{\tn}^{NS}
&=&
\kappa_B\left(
|\,B:+\rangle\!\rangle_{\tn}^{NS}
-|\,B:-\rangle\!\rangle_{\tn}^{NS}
\right), \nn \\
|\,C\rangle\!\rangle_{\tn}^{NS}
&=&\kappa_C\left(
|\,C:+\rangle\!\rangle_{\tn}^{NS}
-|\,C:-\rangle\!\rangle_{\tn}^{NS}
\right), \nn \\
|\,J:\tp\rangle\!\rangle_{\tn}^{NS}
&=&
\frac{\kappa_J}{2}(
|\,J:(+,+),\tp\rangle\!\rangle_{\tn}^{NS}
+|\,J:(-,-),\tp\rangle\!\rangle_{\tn}^{NS}\nn \\
&&
-|\,J:(+,-),\tp\rangle\!\rangle_{\tn}^{NS}
-|\,J:(-,+),\tp\rangle\!\rangle_{\tn}^{NS}),
\end{eqnarray}
where $\kappa_B$, $\kappa_C$ and $\kappa_J$ is the normalization
factor for each states.

The general boundary states can be written as the product of these
states.
The locus of the long string boundary is
\[
 \tl\equiv \tm-1-\tl+\tq \quad \mod{\tn}.
\]
and can be calssified by $\tm$.
When $\tm$ is even, we have two solutions
$\tl=(\tq-1)/2,(\tq-1+\tm)/2$.
In this case, the long string have two loose ends at one boundary
and no loose ends another boundary (Figure 6 left).
When $\tm$ is odd,
we have one solution for each boundary conditions.
In this case, the long string have one loose end at each boundaries
(Figure 6 right).
Thus we will construct the boundary
states for even $\tm$ and odd $\tm$ seperately.

 \begin{figure}[ht]
  \centerline{\epsfxsize=8cm \epsfbox{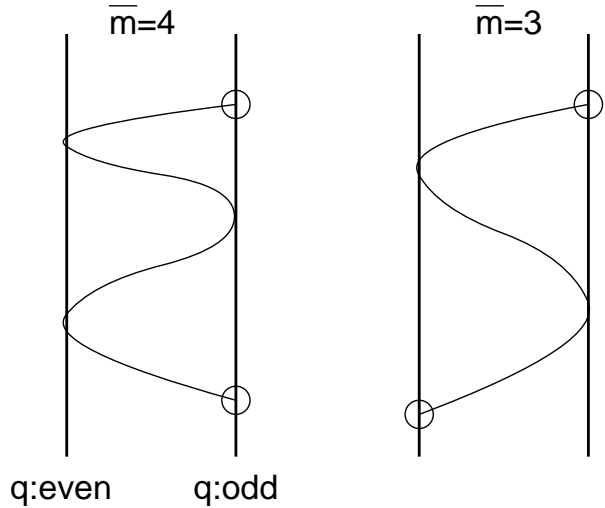}}
  \vskip 3mm
  \caption{Loose ends of long string}
 \end{figure}

For odd $\tm$ case, the long string have one loose end at each
boundaries.
Therefore the boundary and cross-cap states
for the long string is written as,
\footnote{
If the torus amplitude is not considered,
we can use the boundary/cross-cap state which is the products
of the joint states.
}
\begin{eqnarray}
|\,B:f^B\rangle\!\rangle_{\tn}^{NS}
&\equiv&
|\,B\rangle\!\rangle_{\tn}^{NS}
\bigotimes_{i=1}^{(\tm-1)/2}
|\,J(2i-1,2i):\tp_{i}\rangle\!\rangle_{\tn}^{NS},
\\
|\,B:f^C\rangle\!\rangle_{\tn}^{NS}
&\equiv&
|\,C\rangle\!\rangle_{\tn}^{NS}
\bigotimes_{i=1}^{(\tm-1)/2}
|\,J(2i-1,2i):\tp_{i}\rangle\!\rangle_{\tn}^{NS}.
\end{eqnarray}
For these states,
we consider the tadpole cancellation condition as,
\begin{equation}
|\,B:f^B\rangle\!\rangle_{\tn 0}^{NS}-i
|\,B:f^C\rangle\!\rangle_{\tn 0}^{NS}=0.
\end{equation}
This condition can be factorized into the following condition,
\begin{equation}
|\,B\rangle\!\rangle_{\tn}^{NS}-i
|\,C\rangle\!\rangle_{\tn}^{NS}=0.
\label{tadpole cond}
\end{equation}
Thus the tadpole condition is satisfied when,
\begin{equation}
\kappa_B-\kappa_C=0,
\label{tadpole coeff}
\end{equation}
and there is no constraint for $\kappa_J$.

For even $\tm$,
the long string have two loose ends at one boundary
and no loose end at another boundary.
The boundary state $|\,B:f^J\rangle\!\rangle_{\tn}^{NS}$ which have no
loose end does not give any contribution for the tadpole condition.
Therefore, we consider the tadpole condition for the boundary states
which have two loose ends.
We introduce the boundary states as,
\begin{eqnarray}
|\,B:f^A\rangle\!\rangle_{\tn}^{NS}
&\equiv&
|\,B(1)\rangle\!\rangle_{\tn}^{NS}\otimes
|\,B(\tm)\rangle\!\rangle_{\tn}^{NS}
\bigotimes_{i=1}^{\tm/2-1}
|\,J(2i,2i+1):\tp_{i}\rangle\!\rangle_{\tn}^{NS},
\\
|\,B:f^{MS}\rangle\!\rangle_{\tn}^{NS}
&\equiv&
(|\,B(1)\rangle\!\rangle_{\tn}^{NS}\otimes
|\,C(\tm)\rangle\!\rangle_{\tn}^{NS}
+|\,C(1)\rangle\!\rangle_{\tn}^{NS}\otimes
|\,B(\tm)\rangle\!\rangle_{\tn}^{NS}) \nn \\
&&
\bigotimes_{i=1}^{\tm/2-1}
|\,J(2i,2i+1):\tp_{i}\rangle\!\rangle_{\tn}^{NS},
\\
|\,B:f^{KB}\rangle\!\rangle_{\tn}^{NS}
&\equiv&
|\,C(1)\rangle\!\rangle_{\tn}^{NS}\otimes
|\,C(\tm)\rangle\!\rangle_{\tn}^{NS}
\bigotimes_{i=1}^{\tm/2-1}
|\,J(2i,2i+1):\tp_{i}\rangle\!\rangle_{\tn}^{NS}.
\end{eqnarray}
The tadpole cancellation condition is,
\begin{equation}
|\,B:f^{A}\rangle\!\rangle_{\tn 0}^{NS}
+|\,B:f^{MS}\rangle\!\rangle_{\tn 0}^{NS}
+|\,B:f^{KB}\rangle\!\rangle_{\tn 0}^{NS}=0.
\end{equation}
This condition is also factorized
into the condition (\ref{tadpole cond}).
Therefore, if $\kappa_B=\kappa_C$ is satisfied,
the tadpole for $\tm$ long strings is cancelled.

To determine Chan-Paton group, we consider the modular property
for the partition functions calculated before.
Gathering all sectors,
the inner products between the boundary states are written as,
\begin{eqnarray}
{\rm Annulus}&:&\quad
2\kappa_J^{\tm-2}\kappa_{B}^2
\left(
\frac{Z^{0}_0(-\frac{\tm}{\tn\tau})^4}{\eta(-\frac{\tm}{\tn\tau})^8}
-\frac{Z^{0}_1(-\frac{\tm}{\tn\tau})^4}{\eta(-\frac{\tm}{\tn\tau})^8}
\right), \nn \\
\mbox{M\"obius strip}&:&\quad
4i\kappa_J^{\tm-2}\kappa_{B}\kappa_C
\left(
\frac{Z^{0}_0(-\frac{\tm}{\tn\tau}+\frac{1}{2})^4}
{\eta(-\frac{\tm}{\tn\tau}+\frac{1}{2})^8}
-\frac{Z^{0}_1(-\frac{\tm}{\tn\tau}+\frac{1}{2})^4}
{\eta(-\frac{\tm}{\tn\tau}+\frac{1}{2})^8}
\right), \nn \\
{\rm Klein~bottle}&:&\quad
2\kappa_J^{\tm-2}\kappa_{C}^2
\left(
\frac{Z^{0}_0(-\frac{\tm}{\tn\tau})^4}{\eta(-\frac{\tm}{\tn\tau})^8}
-\frac{Z^{0}_1(-\frac{\tm}{\tn\tau})^4}{\eta(-\frac{\tm}{\tn\tau})^8}
\right).
\end{eqnarray}
After the modular transformation, the amplitudes are written as,
\begin{eqnarray}
{\rm Annulus}&:&\quad
2\kappa_J^{\tm-2}\kappa_{B}^2
\left(
\frac{2\tk \tau}{i\tm}
\right)^{-4}
\left(
\frac{Z^{0}_0(\frac{2\tk\tau}{\tm})^4}{\eta(\frac{2\tk\tau}{\tm})^8}
-\frac{Z^{1}_0(\frac{2\tk\tau}{\tm})^4}{\eta(\frac{2\tk\tau}{\tm})^8}
\right), \nn\\
\mbox{M\"obius strip}&:&\quad
4\kappa_J^{\tm-2}\kappa_{B}\kappa_C
\left(
\frac{\tk \tau}{i\tm}
\right)^{-4}
\left(
\frac{Z^{0}_0(\frac{\tk\tau}{2\tm}+\frac{1}{2})^4}
{\eta(\frac{\tk\tau}{2\tm}+\frac{1}{2})^8}
-\frac{Z^{0}_1(\frac{\tk\tau}{2\tm}+\frac{1}{2})^4}
{\eta(\frac{\tk\tau}{2\tm}+\frac{1}{2})^8}
\right), \nn\\
{\rm Klein~bottle}&:&\quad
2\kappa_J^{\tm-2}\kappa_{C}^2
\left(
\frac{2\tk \tau}{i\tm}
\right)^{-4}
\left(
\frac{Z^{0}_0(\frac{2\tk\tau}{\tm})^4}{\eta(\frac{2\tk\tau}{\tm})^8}
-\frac{Z^{1}_0(\frac{2\tk\tau}{\tm})^4}{\eta(\frac{2\tk\tau}{\tm})^8}
\right),\label{modular part fun}
\end{eqnarray}
where we denoted $\tk=\tn/2$.
These expressions should be compared with the partition functions which
calculated in the open string sector as,
\begin{eqnarray}
{\rm Annulus}&:&\quad
\frac{N^2}{4}
\left(
\frac{2m \tau}{in}
\right)^{-4}
\left(
a\frac{Z^{0}_0(\frac{m\tau}{n})^4}{\eta(\frac{m\tau}{n})^8}
+c\frac{Z^{1}_0(\frac{m\tau}{n})^4}{\eta(\frac{m\tau}{n})^8}
\right),\nn \\
\mbox{M\"obius strip}&:&\quad
\eta N
\left(
\frac{m\tau}{ik}
\right)^{-4}
\left(
a\frac{Z^{0}_0(\frac{m\tau}{2k}+\frac{1}{2})^4}
{\eta(\frac{m\tau}{2k}+\frac{1}{2})^8}
-b\frac{Z^{0}_1(\frac{m\tau}{2k}+\frac{1}{2})^4}
{\eta(\frac{m\tau}{2k}+\frac{1}{2})^8}
\right), \nn \\
{\rm Klein~bottle}&:&\quad
\frac{1}{2}
\left(
\frac{k \tau}{in}
\right)^{-4}
\left(
\frac{Z^{0}_0(\frac{2k\tau}{n})^4}{\eta(\frac{2k\tau}{n})^8}
-\frac{Z^{1}_0(\frac{2k\tau}{n})^4}{\eta(\frac{2k\tau}{n})^8}
\right),\label{open part fcn}
\end{eqnarray}
where $N$ is the rank of the Chan-Paton group and $\eta=\pm 1$ comes
from the action of $\Omega$ on the Chan-Paton group.
When $\eta=1$, this group is $Sp(N)$ and
When $\eta=-1$, this group is $SO(N)$.
Since $\kappa_J$ should not depend on the length factor,
we need to impose $\kappa_J=1$.

By comparing (\ref{modular part fun}) and (\ref{open part fcn}),
we need to impose,
\begin{itemize}
\item Annulus:  $m=2\tk$, $n=\tm$

\item M\"obius strip:  $m=\tk$, $k=\tm$

\item Klein bottle:  $k=\tk$, $n=\tn$

\end{itemize}
Under these correspondences and tadpole condition (\ref{tadpole coeff}),
we obtain the following relations,
\begin{eqnarray}
&&
\frac{aN^2}{4}=2^9\kappa_B^2, \quad
\frac{cN^2}{4}=-2^9\kappa_B^2, \nn \\
&&
aN\eta=-2^6\kappa_B^2, \quad
-bN\eta=-2^6\kappa_B^2, \nn \\
&&
2\kappa_B^2=\frac{1}{4}.
\end{eqnarray}
These equations can be solved as,
\begin{eqnarray}
&&
a=b=-c=\frac{1}{2},\quad \kappa_B=\frac{1}{2}, \nn \\
&&
\eta=-1, \quad N=2^5.
\label{sloution for tadpole}
\end{eqnarray}
Thus the consistent Chan-Paton group is $SO(32)$, and this is the
standard gauge group for Type I theory.

For R-$\widetilde{{\rm R}}$ sector,
the tadpole cancellation condition can be
found in similar way.
The GSO-projected boundary and cross-cap states for
R-$\widetilde{{\rm R}}$ sector is written as,
\begin{eqnarray}
|\,B\rangle\!\rangle_{\tn}^{R}
&=&
i\kappa_B^{\prime}\left(
|\,B:+\rangle\!\rangle_{\tn}^{R}
+|\,B:-\rangle\!\rangle_{\tn}^{R}
\right), \nn \\
|\,C\rangle\!\rangle_{\tn}^{R}
&=&
i\kappa_C^{\prime}\left(
|\,C:+\rangle\!\rangle_{\tn}^{R}
+|\,C:-\rangle\!\rangle_{\tn}^{R}
\right), \nn \\
|\,J:\tp\rangle\!\rangle_{\tn}^{R}
&=&
\frac{\kappa_J^{\prime}}{2}(
|\,J:(+,+),\tp\rangle\!\rangle_{\tn}^{R}
+|\,J:(-,-),\tp\rangle\!\rangle_{\tn}^{R}\nn \\
&&
+|\,J:(+,-),\tp\rangle\!\rangle_{\tn}^{R}
+|\,J:(-,+),\tp\rangle\!\rangle_{\tn}^{R}),
\end{eqnarray}
Since the coefficient $\kappa_{J}^{\prime}$ should not depend on the
length of the long string, we set $\kappa_{J}^{\prime}=1$
The boundary state for odd $\tm$ is expressed as the products of the
boundary, cross-cap and joint states,
\begin{eqnarray}
|\,B:f^B\rangle\!\rangle_{\tn}^{R}
&\equiv&
|\,B\rangle\!\rangle_{\tn}^{R}
\bigotimes_{i=1}^{(\tm-1)/2}
|\,J(2i-1,2i):\tp_{i}\rangle\!\rangle_{\tn}^{R},
\\
|\,B:f^C\rangle\!\rangle_{\tn}^{R}
&\equiv&
|\,C\rangle\!\rangle_{\tn}^{R}
\bigotimes_{i=1}^{(\tm-1)/2}
|\,J(2i-1,2i):\tp_{i}\rangle\!\rangle_{\tn}^{R}.
\end{eqnarray}
The boundary state for even $\tm$ can be written
as NS-$\widetilde{{\rm NS}}$ sector.
The tadpole condition is also factorized into
that of the irreducible combination.

The non-zero inner products between these boundary states are written as,
\begin{eqnarray}
{\rm Annulus} &:&\quad
-(\kappa_B^{\prime})^2\frac{Z^{1}_{0}(-\frac{\tm}{\tn\tau})^4}
{\eta(-\frac{\tm}{\tn\tau})^8}, \nn \\
\mbox{M\"obius strip}&:&\quad
-\kappa_B^{\prime}\kappa_C^{\prime}
\frac{Z^{1}_{0}(-\frac{\tm}{\tn\tau}+\frac{1}{2})^4}
{\eta(-\frac{\tm}{\tn\tau}+\frac{1}{2})^8}, \nn \\
{\rm Klein bottle} &:&\quad
-(\kappa_C^{\prime})^2\frac{Z^{1}_{0}(-\frac{\tm}{\tn\tau})^4}
{\eta(-\frac{\tm}{\tn\tau})^8},
\end{eqnarray}
where the minus signs come from the ghost sector.
\footnote{
In this section, we are considering the superstring theory 
on $\bR^{1,1}\times S^{N}\bR^8$.}
These inner products should be compared with the partition function
which calculates in the open string sector,
\begin{eqnarray}
{\rm Annulus}&:&\quad
-N^2\frac{b}{4}\left(
\frac{2m\tau}{in}
\right)^{-4}
\frac{Z^{0}_{1}(\frac{m\tau}{n})^4}{\eta(\frac{m\tau}{n})^8},
\nn \\
\mbox{M\"obius strip}&:&\quad
c\eta N\left(
\frac{m\tau}{ik}
\right)^{-4}
\frac{Z^{1}_{0}(\frac{m\tau}{2k}+\frac{1}{2})^4}
{\eta(\frac{m\tau}{2k}+\frac{1}{2})^8},
\nn \\
{\rm Klein bottle} &:&\quad
-\frac{1}{2}\left(
\frac{2k\tau}{in}
\right)^{-4}
\frac{Z^{0}_{1}(\frac{2k\tau}{n})^4}{\eta(\frac{2k\tau}{n})^8}.
\end{eqnarray}
We can determine the unknown factors as (\ref{sloution for tadpole}) and
$\kappa_B^{\prime}=-\kappa_C^{\prime}$.
Thus the tadpole cancellation condition can be satistied for
NS-$\widetilde{{\rm NS}}$ and R-$\widetilde{{\rm R}}$ consistently 
for the Chan-Paton gauge group $SO(32)$.

\section{Discussion}

In this paper, we showed that the open superstring
on symmetric product can be interpreted
as the second quantized Type I superstring theory.
From the calculation of the partition function,
we found that the oriented open string bits are connected to
become the unoriented open/closed strings.
From the construction of the boundary states,
we found that they are classified into three types of the
boundary states and explain the change of the world-sheet topology.
Although we mainly discussed about Type I superstring theory,
the consideration for the Type 0 superstring theory
can be done in similar way by taking diagonal GSO projection.
Such a theory expresses the second-quantized Type 0 string theory.

In our model, the description of the D-brane can be done straightforwardly
by changing the loose ends of the long strings.
Furthermore, if one want to treat the anti-D-brane, we should make
the opposite GSO projection for each boundary/cross-cap/joint state.
As an application of our model, we hope that our description of
the second-quantized superstring theory becomes useful for the
calculation of the tachyon condensation in the string field theory.\cite{SEN-ZWIEBACH,BSZ}

Here we comment about the interaction vertex operator.
One of the advantage of the representation of the string field theory
by the symmetric product, is the economical description of the
interactions.
It was verified that 
the interaction vertex operator is consistently introduced to the 
matrix string theory by evaluating the four-point functions.\cite{Frolov}\cite{AAF}
By evaluating the four-tachyon scattering amplitude in the large $N$
limit as \cite{Frolov}, 
it is verified that three-open string interaction is consistently
introduced to the open bosonic string theory on the symmetric product 
$S^N\bR^{24}$ 
as the twist operator with conformal weight $\frac{3}{2}$ 
on the boundary of the world-sheet.\cite{PHD}
As already discussed by C.V.Johnson\cite{Johson}, there is only one
open superstring vertex operator $\Sigma\cdot\Phi_{KL}$.
The operator $\Phi_{KL}$ interchanges $K$'th and $L$'th open strings
at the boundary.
The operator $\Sigma$ expresses the spin field and
this is the picture-changing operator in NSR formalism\cite{DVV}.
Naively, in terms of the boundary states, this operator mixes
the boundary/cross-cap states and the joint states as,
\begin{eqnarray}
\bullet
&&
\be{\frac{\alpha(\phi_K+\phi_L)}{2}}
|\,J(KL):\alpha,(\be{\phi_K/2},\be{\phi_L/2})
\rangle\!\rangle
\nn \\
&\leftrightarrow&
\be{\frac{\alpha\phi_K^{\prime}}{2}}
|\,B(K):\alpha,\be{\phi_K^{\prime}/2}\rangle\!\rangle \otimes
\be{\frac{\alpha\phi_L^{\prime}}{2}}
|\,B(L):\alpha,\be{\phi_L^{\prime}/2}\pm\rangle\!\rangle
\nn \\
\bullet
&&
\be{\frac{\alpha(\phi_K+\phi_M)}{2}}
|\,J(KM):\alpha,(\be{\phi_K/2},\be{\phi_M/2})
\rangle\!\rangle
\nn \\
&&
\otimes
\be{\frac{\alpha(\phi_L+\phi_N)}{2}}
|\,J(LN):\alpha,(\be{\phi_L/2},\be{\phi_N/2})
\rangle\!\rangle
\nn \\
&\leftrightarrow&
\be{\frac{\alpha(\phi_L^{\prime}+\phi_M^{\prime})}{2}}
|\,J(LM):\alpha,(\be{\phi_L^{\prime}/2},\be{\phi_M^{\prime}/2})
\rangle\!\rangle
\nn \\
&&
\otimes
\be{\frac{\alpha(\phi_K^{\prime}+\phi_N^{\prime})}{2}}
|\,J(KN):\alpha,(\be{\phi_K^{\prime}/2},\be{\phi_N^{\prime}/2})
\rangle\!\rangle.
\end{eqnarray}
We depicted above interaction in Figure 7.
However we do not know
whether this interaction vertex becomes Lorentz invariant and
can be consistently introduced to our model.
This is one of the most important issue which should be clarified in the
future work.

 \begin{figure}[ht]
  \centerline{\epsfxsize=10cm \epsfbox{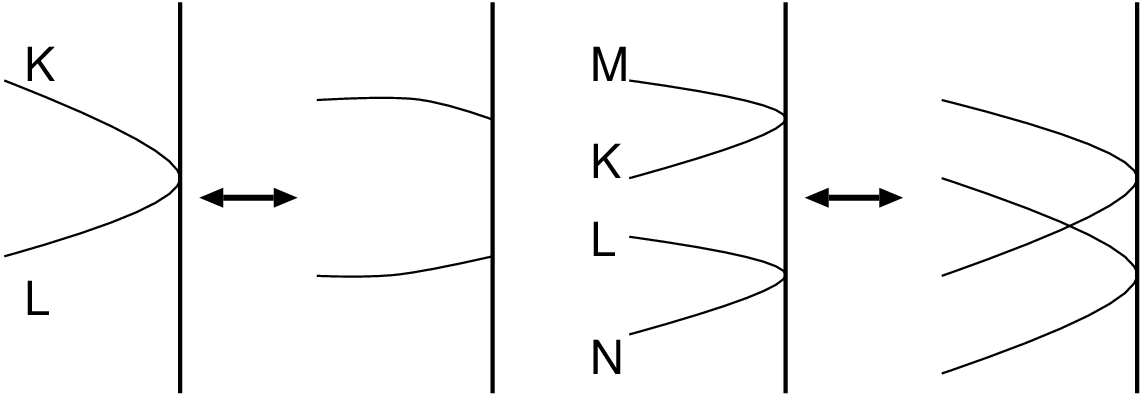}}
  \vskip 3mm
  \caption{Open string interaction}
 \end{figure}

The intereting problem for the future work is as follows.
The description of the second-quantized superstring theory is
discussed in a slightly different way.\cite{discrete}
The fermionic $\bZ_2$-twisting is identified with the $H^2(S_N,U(1))$,
and the supersymmetry is described in more compact way.
The consideration for the open string in this context will be
interesting.
The other direction is the non-trivial background.
Recently the string theory in the non-commutative
geometry has been developing.\cite{SW}
The Matrix string theory in the B-field background is also discussed.\cite{Semenoff}
Whether the open string theory on the symmetric product of the
non-commutative geomtry naturally describes the string field theory
in the non-commutative geometry is very interesting issue.
Furthermore, the matrix model is discussed in the Melvin background.\cite{Melvin}
The extension of our model to this background will be also interesting.
The string theory on the symmetric product of
the curved background as Calabi-Yau manifold was partially
discussed.\cite{Klemm}
We expect that the construction of the boundary states which represent
the D-branes wrapping around the supersymmetric cycle 
or holomorphic cycle 
may shed new light on the string field theory on the Calabi-Yau
manifolds\cite{CS,Berkovits CY} and
the open string instanton calculations.\cite{Kawai}

\vskip 5mm
\noindent{\bf Acknowledgement}:
The author is obliged to Y.Matsuo for
his useful suggestions.

The author is supported by the JSPS fellowship for Young Scientist.


\begin{thebibliography}{999}
 \bibitem{Matsuo-Fuji}
  H.Fuji and Y.Matsuo, ``Open String on Symmetric Product'', \\
   Int.J.Mod.Phys.A{\bf 16} (2001) 557-608, hep-th/0005111.
 \bibitem{Kaku-Kikkawa}
  M.Kaku and K.Kikkawa, 
    ``Field Theory of Relativistic Strings. I. Trees'',
    Phys.Rev.D{\bf 10} (1974) 1110-1133; 
    ``Field Theory of Relativistic Strings. II. Loops and Pomerons'',
    Phys.Rev.D{\bf  10} (1974) 1823-1843.  
 \bibitem{Cremmer-Gervais}
  E.Cremmer and J.L.Gervais,
  ``Combining and Splitting Relativistic Strings'',
   Nucl.Phys.B{\rm 76} (1974) 209-230;
  ``Infinite Component Field Theory of Interacting Relativistic Strings
    and Dual Theory'', 
    Nucl.Phys.B{\rm 90} (1975) 410-460.
 \bibitem{WITTEN}
    E.Witten, 
     ``Non-commutative Geometry and String Field Theory'',
     Nucl.Phys.B{\bf 268} (1986) 79-324;\\
     ``Interacting Field Theory of Open Superstring Field Theory'',\\
     Nucl.Phys.B{\bf 276} (1986) 291-324.
 \bibitem{HIKKO}
   H.Hata, K.Itoh, T.Kugo, H.Kunitomo and K.Ogawa, 
    ``Covariant String Field Theory'', 
    Phys.Rev.D{\bf 34} (1986) 2360-2429;
    ``Covariant String Field Theory, II'', 
    Phys.Rev.D{\bf 35} (1987) 1318-1355.
 \bibitem{Mandelstam}
  S.Mandelstam,
  ``Interacting-String Picture of the Neveu-Schwarz-Ramond Model'', 
    Nucl.Phys.B{\bf 69} (1974) 77-106;
  ``Lorentz Properties of The Three-String Vertex'',
    Nucl.Phys.B{\bf 83} (1974) 413-439.
 \bibitem{Green}
  M.B.Green and J.H.Schwarz, 
  ``Superstring Interactions'', Nucl.Phys.B{\bf 218} (1983) 43-88;
  ``Superstring Field Theory'', Nucl.Phys.B{\bf 243} (1984) 475-536.
  \bibitem{Berkovits}
  N.Berkovits,
   ``Super-Poincare Invariant Superstring Field Theory'',
   Nucl.Phys.B{\bf 450} (1995) 90-102, hep-th/9503099.
 \bibitem{BFSS}
  T.Banks, W.Fischler, S.H.Shenker and L.Susskind, ``M Theory As A Matrix 
   Model:A Conjecture'',Phys.Rev.D{\bf 55} (1997) 5112-5128, hep-th/9610043.
 \bibitem{DVV}
  R.Dijkgraaf, E.Verlinde and H.Verlinde, ``Matrix String Theory'',
   Nucl.Phys.B{\bf 500} (1997) 43-61, hep-th/9703030.
 \bibitem{MOTL} 
   L.Motl, 
    ``Proposals on Nonperturbative Superstring Interactions'',
     hep-th/9701025.
 \bibitem{BANKS-SEIBERG}
   T.Banks and N.Seiberg, 
    ``Strings from Matrices'', 
     Nucl. Phys. {\bf B497} (1997) 44-55, hep-th/9702187.
 \bibitem{DMVV}
  R.Dijkgraaf, G.Moore, E.Verlinde and H.Verlinde, 
  ``Elliptic Genera of Symmetric Products and Second Quantized Strings'',
    Commun.Math.Phys. {\bf 185} (1997) 197-209, hep-th/9608096.
 \bibitem{Susskind}
  L.Susskind, ``Another Conjecture about M(atrix) Theory'', hep-th/9704080,
 \bibitem{Bantay}
  P.Bantay, ``Characters and modular properties of permutation
 orbifolds'', Phys.Lett.B{\bf 419} (1998) 175-178, hep-th/9708120;
 ``Permutation orbifolds'', hep-th/9910079;
    Z.Kadar,
    ``The torus and the Klein Bottle amplitude of permutation
     orbifolds'',
   Phys.Lett.B{\bf 484} (2000) 289-294, 
     hep-th/0004122.
 \bibitem{Vanhove}
  I.Kostov and P.Vanhove, 
  ``Matrix String Partition Functions'', 
   Phys.Lett.B{\bf 444} (1998) 196-203, hep-th/9809130; 
   C.Bachas, C.Fabre, E.Kiritsis, N.A.Obers, and P.Vanhove, 
   ``Heterotic / type I duality and D-brane instantons'', 
   Nucl.Phys.B{\bf 509} (1998) 33-52, hep-th/9707126.
 \bibitem{sugino0}
  F.Sugino,
  ``Cohomological Field Theory Approach to Matrix Strings''
   Int.J.Mod.Phys.A{\bf 14} (1999) 3979-4002, hep-th/9904122.
 \bibitem{Semenoff}
  G.Griani, M.Orsel and Semenoff,
   ``Matrix strings in a B-field'',JHEP.{\bf 0107} (2001) 004, hep-th/0104112.
 \bibitem{Smolin}
 L.Smolin, 
 ``M theory as a matrix extension of Chern-Simons theory'', 
    Nucl.Phys.B{\bf 591} (2000) 227-242;
 ``The cubic matrix model and a duality between strings and loops'',
    hep-th/0006137;
 ``The exceptional Jordan algebra and the matrix string'',
    hep-th/0104050.
 \bibitem{Sugino}
   F.Sugino and P.Vanhove,
   `` U-duality from Matrix Membrane Partition Function'', 
   Phys.Lett.B{\bf 522} (2001) 145-154, hep-th/0107145.
 \bibitem{Sekino-Yoneya}
  Y.Sekino and T.Yoneya,
   ``From Supermembrane to Matrix String'',
     Nucl.Phys.B{\bf 619} (2001)22-50, hep-th/108176.
 \bibitem{Johson}
  Clifford V.Johnson, ``On Second-Quantized Open Superstring Theory'',\\
    Nucl.Phys.B{\bf 537} (1999) 144-160, hep-th/9806115.
 \bibitem{Polchinski}
  J.Polchinski, ``String Theory I,II'', Cambridge University Press (1998). 
 \bibitem{Lang}
  S.Lang, ``Introduction to Modular Forms '',
  Graduate Texts in Mathematics {\bf 222}, Springer-Verlarg (1976).
 \bibitem{SERRE}
   J.-P.Serre, 
    ``A Course in Arithmetics'',
    Graduate Texts in Mathematics {\bf 7}, Springer-Verlag (1973).
 \bibitem{apostol}
  T.M.Apostol, ``Modular Functions and Dirichlet Series in Number Theory'',
   Graduate Texts in Mathematics {\bf 41},
   Springer-Verlarg (1976).
 \bibitem{Pradisi-Sagnotti}
 G.Pradisi and A.Sagnotti, ``Open String Orbifolds'',
   Phys.Lett.B{\bf 216} (1989) 59-67.
 \bibitem{Ishibashi Ph.D}
  N.Ishibahi and T.Onogi, ``Open String Model Building'', 
  Nucl.Phys.B{\bf  318} (1989) 239.
 \bibitem{Harvey-Minahan}
 J.A.Harvey and J.A.Minahan, ``Open Strings on Orbifolds'',
  Phys.Lett.B{\bf 188} (1987) 44-50.
 \bibitem{Cai-Polchinski}
  Y.Cai and J.Polchinski, ``Consistency of the Open Super String'', 
   Nucl.Phys.B{\bf 296} (1988) 91-128.
 \bibitem{Frolov}
    G.E.Arutyunov and S.A.Frolov,
 ``Virasoro amplitude from the
 $S^N\bR^{24}$ orbifold sigma model'', 
   Theor.Math.Phys.{\bf 114} (1998) 43-66, hep-th/9708129.
 \bibitem{AAF}
 G.E.Arutyunov and S.A.Frolov,
 ``Four graviton scattering amplitude from $S^N{\bf R}^{8}$
    supersymmetric orbifold sigma model'', 
  Nucl.Phys.B{\bf 524} (1998) 159-206, hep-th/9712061;
  G.Arutyunov, S.Frolov and  A.Polishchuk
 ``On Lorentz invariance and supersymmetry of 
  four particle scattering amplitudes
   in $S^N\bR^8$ orbifold sigma model'', 
  Phys.Rev.D{\bf 60} (1999) 066003, hep-th/9812119.
 \bibitem{PHD}
 H.Fuji, Ph.D thesis in University of Tokyo, 2002.
 \bibitem{SEN-ZWIEBACH}
  A. Sen and B. Zwiebach, 
  ``Tachyon condensation in string field theory''
  JHEP{\bf 0003} (2000) 002, hep-th/9912249.
 \bibitem{BSZ}
  N. Berkovits, A. Sen and B. Zwiebach,
  ``Tachyon condensation in superstring field theory'', 
  Nucl.Phys.B{\bf 587} (2000) 147-178.
  hep-th/0002211.
 \bibitem{discrete}
  R.Dijkgraaf, ``Discrete Torsion and Symmetric Products'', hep-th/9912101.
 \bibitem{SW}
 N.Seiberg and E.Witten, ``String Theory and Noncommutative Geometry'',
  JHEP {\bf 9909} (1999) 032, hep-th/9908142.
 \bibitem{Melvin}
 L.Motl, ``Melvin Matrix Models'', hep-th/0107002.
 \bibitem{Klemm}
  A.Klemm and M.G.Schmid,
  ``Orbifolds by Cyclic Permutations of Tensor Product
   Conformal Field Theories'', 
   Phys.Lett.B{\bf 245} (1990) 53-58;
  J.Fuch, A.Klemm and M.G.Schmid,
  ``Orbifolds by Cyclic Permutations in Gepner Type
   Superstrings and in the Corresponding Calabi-Yau
   Manifolds'',
  Annals. Phys.{\bf 214} (1992) 221-257.
 \bibitem{CS}
  E.Witten, ``Chern-Simons Theory As A String Theory'', hep-th/9207094.
 \bibitem{Berkovits CY}
  N.Berkovits,
  ``Covariant Quantization of the Green-Schwarz Superstring in a Calabi-Yau
   Background'',
  Nucl.Phys.B{\bf 431} (1994) 258-272, hep-th/9404162;
  ``Review of Open Superstring Field Theory'', 
  hep-th/0105230;
 ``The Ramond Sector of Open Superstring Field Theory'', 
  hep-th/0109100.
\bibitem{Kawai}
  T.Kawai and K.Yoshioka,
 ``String Partition Functions and Infinite Products'', \\
  Adv.Theor.Math.Phys. {\bf 4} (2001) 397-485, hep-th/0002169.
\end{thebibliography}
\end{document}